\documentclass[numbers,webpdf,imaiai]{ima-authoring-template}%
\usepackage{graphicx}
\usepackage{subfigure,wrapfig,todonotes}
\usepackage{cancel}
\usepackage{tabularx}
\usepackage{booktabs} 

\graphicspath{{Fig/}}

\theoremstyle{thmstyletwo}%
\numberwithin{equation}{section}

\newcommand{\xslnote}[1]{{\color{blue}[ #1 -- SHERRY]}}
\newcommand{\deknote}[1]{{\color{black}#1}}
\newcommand{\wbnote}[1]{{\color{black}#1}}

\newcommand{\ylnote}[1]{{\color{black}#1}}

\newlength{\doublecolumnimgwidth}
\newlength{\singlecolumnimgwidth}
\usepackage{amsopn}

\newcommand{\ignore}[1]{}
\newcommand{\hh}{$\mathcal{H}^2$ }

\algblockdefx{ParallelFor}{EndParallelFor}[1]{\textbf{parallel for} #1 \textbf{do}}{\textbf{end for}}

\begin{document}

\DOI{DOI}
\copyrightyear{2025}
\vol{00}
\pubyear{2026}
\access{Advance Access Publication Date: Day Month Year}
\appnotes{Paper}
\copyrightstatement{Published by Oxford University Press on behalf of the Institute of Mathematics and its Applications. All rights reserved.}
\firstpage{1}


\title[Linear Complexity $H^2$ Solver]{\Large Linear Complexity $\mathcal{H}^2$ Direct Solver for Fine-Grained Parallel Architectures}

\author{Wajih Boukaram*\ORCID{0000-0002-2958-9344}
\address{\orgdiv{Applied Mathematics and Computational Sciences}, \orgname{King Abdullah University of Science and Technology}, \orgaddress{\street{4700 KAUST}, \postcode{23955-6900}, \state{Thuwal}, \country{KSA}}}}
\author{David Keyes\ORCID{0000-0002-4052-7224}
\address{\orgdiv{Applied Mathematics and Computational Sciences}, \orgname{King Abdullah University of Science and Technology}, \orgaddress{\street{4700 KAUST}, \postcode{23955-6900}, \state{Thuwal}, \country{KSA}}}}
\author{Sherry Li\ORCID{0000-0002-0747-698X}
\address{\orgdiv{Applied Mathematics and Computational Research Division}, \orgname{Lawrence Berkeley National Laboratory}, \orgaddress{\street{1 Cyclotron Road, MS 50A-3111}, \postcode{94710}, \state{Berkeley, California}, \country{USA}}}}
\author{Yang Liu\ORCID{0000-0003-3750-1178}
\address{\orgdiv{Applied Mathematics and Computational Research Division}, \orgname{Lawrence Berkeley National Laboratory},  \orgaddress{\street{1 Cyclotron Road, MS 50A-3111}, \postcode{94710}, \state{Berkeley, California}, \country{USA}}}}
\author{George Turkiyyah\ORCID{000-0002-1692-5812}
\address{\orgdiv{Applied Mathematics and Computational Sciences}, \orgname{King Abdullah University of Science and Technology}, \orgaddress{\street{4700 KAUST}, \postcode{23955-6900}, \state{Thuwal}, \country{KSA}}}}

\authormark{Boukaram, et al.}

\corresp[*]{Corresponding author: \href{email:wajih.boukaram@kaust.edu.sa}{wajih.boukaram@kaust.edu.sa}}

\received{2}{9}{2025}


\abstract{
We present factorization and solution phases for a new linear complexity direct solver designed for concurrent batch operations on fine-grained parallel architectures, for matrices amenable to hierarchical representation. We focus on the strong-admissibility-based $\mathcal{H}^2$ format, where strong recursive skeletonization factorization compresses remote interactions. We build upon previous implementations of $\mathcal{H}^2$ matrix construction for efficient factorization and solution algorithm design, which are illustrated graphically in stepwise detail. \deknote{The algorithms are ``blackbox'' in the sense that the only inputs are the matrix and right-hand side, without analytical or geometrical information about the origin of the system.} 
We demonstrate linear complexity scaling in both time and memory on four representative families of dense matrices up to one million in size. Parallel scaling up to 16 threads is enabled by a multi-level matrix graph coloring and avoidance of dynamic memory allocations thanks to prefix-sum memory management. An experimental backward error analysis is included. We break down the timings of different phases, identify phases that are memory-bandwidth limited, and discuss alternatives for phases that may be sensitive to the trend to employ lower precisions for performance. 
}



\maketitle

\pagebreak

\section{Introduction}


Hierarchically off-diagonal low-rank structure has been used to develop fast
algebraic solvers for many systems, including various structured matrices, e.g.,
Toeplitz and Cauchy~\cite{chandrasekaran2008superfast}, as well as broader problems,
such as integral equation methods for acoustics,
electromagnetics~\cite{hackbusch1999sparse,bebendorf2003existence,borm2003introduction},
differential equation-based PDE solvers~\cite{xia2013randomized,ghysels2017robust},
machine learning methods like kernel ridge regression~\cite{chavez2020scalable}
and Gaussian processes~\cite{ambikasaran2015fast}.

There is a diverse family of hierarchical matrix techniques, including the $\mathcal{H}$/$\mathcal{H}^2$ format~\cite{hackbusch1999sparse,bebendorf2003existence,borm2003introduction}, the hierarchically off-diagonal low-rank format (HODLR)~\cite{ambikasaran2013mathcal}, the hierarchically semi-separable format (HSS)~\cite{chandrasekaran2007fast} or hierarchically block separable format (HBS)~\cite{gillman2012direct}, the inverse fast multipole method (IFMM) \cite{coulier2017inverse} and the hierarchical interpolative factorization (HIF) algorithms \cite{l2016hierarchical}. 
These formats can be characterized by the so-called admissibility condition, which determines how much separated interactions can be low-rank compressed. The optimal choice of hierarchical format depends on the particular application, including the dimensionality and discretization scheme. 
For high-dimensional problems, weak-admissibility-based formats such as HODLR, HSS, HBS typically cannot achieve quasi-linear complexities with the exception of HIF, which, however, has been demonstrated only with regular grid-based discretization. On the other hand, strong-admissibility-based formats can attain quasi-linear (e.g. $\mathcal{H}$) and linear complexities (e.g., $\mathcal{H}^2$ and IFMM) for high-dimensional problems.
This paper addresses the strong-admissibility-based $\mathcal{H}^2$ format.

In earlier work~\cite{boukaram2025}, we developed a novel linear-complexity bottom-up
sketching-based algorithm for constructing an $\mathcal{H}^2$ matrix, and
presented its high performance GPU implementation.
This paper builds upon that work and further develops efficient factorization
and solution algorithms likewise suitable for fine-grained architectures  \cite{keyes20_haha}.
It is worth mentioning that once the $\mathcal{H}^2$ matrix has been constructed, efficient (i.e., low-prefactor) inversion algorithms have also been recently developed \cite{minden2017recursive,Ma2019} and parallelized \cite{liang2024on,ma2022scalable}.
Our algorithms thereby fill a gap in GPU computing and GPU code of the algorithm presented herein is in progress, to follow the CPU implementation.

\subsection{Background on Hierarchical Matrices}

Hierarchical matrices offer a compact and efficient way to represent dense matrices that possess data sparsity, where specific sub-blocks can be accurately approximated using low-rank structures. Various formulations of hierarchical matrices have emerged, distinguished primarily by their strategies for block partitioning and the methods used to construct and store the basis vectors for low-rank approximations. The block partitioning process typically begins by hierarchically clustering the indices of a matrix $A$ into a cluster tree $I$. A dual tree traversal is then performed on the cluster tree, generating pairs of clusters $(s, t)$ that are evaluated against an admissibility condition. This condition determines whether the matrix block corresponding to the cluster pair can be effectively approximated by a low-rank matrix. In this work, we adopt the general admissibility condition, denoted as \texttt{adm}, which assesses the compressibility of a block based on the distance \texttt{Dist} between the bounding boxes of $s$ and $t$, and the average of their diameters $D$:
\begin{equation}
\texttt{adm}(s,t)=1,~{\rm{if}}~\frac{D(s) + D(t)}{2} \leq \eta \cdot \texttt{Dist}(s, t)\label{eqn:admissibility}
\end{equation}
The parameter $\eta$ controls the strength of the admissibility condition, starting from what is usually referred to as strong admissibility for values of $\eta \geq 0.5$ and weakening as it is increased, coarsening the matrix structure. The general admissibility condition is used to guide a dual tree traversal, which constructs a matrix tree in which each node represents a cluster pair $(s, t)$ at the same level of the cluster tree. If a cluster pair fails to satisfy the admissibility criterion, the traversal proceeds recursively on the four child pairs derived from $s$ and $t$. This process continues until the corresponding matrix block becomes small enough to store in its original dense form. The complete set of leaf nodes in the resulting matrix tree defines the final block partitioning of the matrix.

\begin{wrapfigure}[18]{l}{0.45\textwidth}
  \centering
  \vspace{-10pt}
  \includegraphics[width=0.9\linewidth]{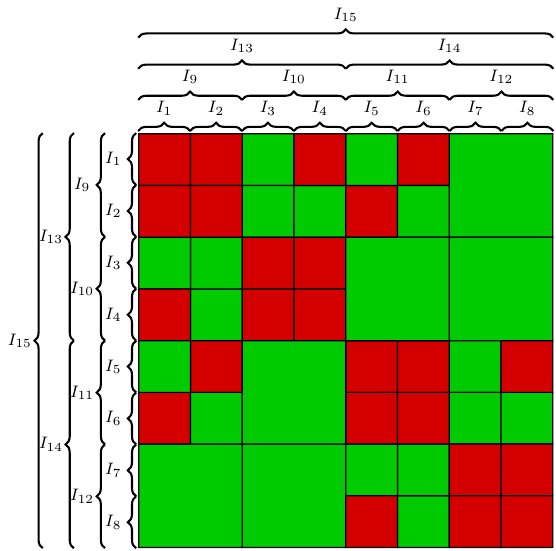}  
  \caption{A hierarchically partitioned  matrix and the cluster tree $I$.}
\label{fig:original_hmatrix_with_cluster_tree}
\end{wrapfigure}

\begin{figure}[b]
  \centering
  \vspace{-15pt}
\includegraphics[width=0.7\linewidth]{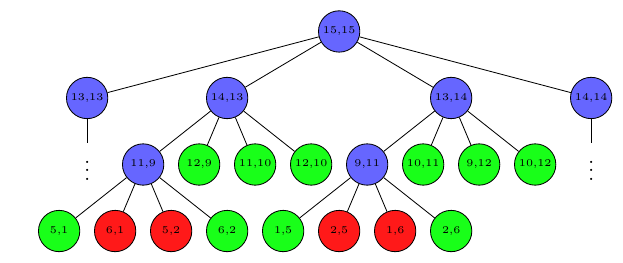}  
  \caption{The matrix tree for the hierarchical matrix in Fig. \ref{fig:original_hmatrix_with_cluster_tree} representing the inadmissible blocks in blue, the admissible leaves in green and the inadmissible leaves in red. In general, the matrix tree is not a complete tree. The ellipses represent complete subtrees, omitted for brevity.}
  \label{fig:matrix_tree}
\end{figure}

Figure~\ref{fig:original_hmatrix_with_cluster_tree} shows the hierarchically partitioned matrix $A$ and its row and column indices hierarchically clustered into a cluster tree $I$. 
This block partitioning of the matrix is formed from the leaves of the hierarchical matrix tree of Fig.~\ref{fig:matrix_tree}, generated from a dual tree traversal on the cluster tree. 
The complete subtrees $A_{13,13}$ and $A_{14,14}$ produced from the diagonal blocks are omitted for brevity. The inadmissible leaves of the tree are shown in red, the inadmissible inner nodes are shown in blue, and admissible leaves are shown in green.
All the inadmissible nodes
can be interpreted as forming a block sparse matrix. A key property of hierarchical matrices, ultimately leading to log-linear or linear complexity estimates for storage and factorization and solution operations, is that the number of non-zero blocks 
in any row or column of such a block sparse matrix is bounded by a constant that remains independent of the overall problem size. This constant is referred to as the \emph{sparsity constant} $C_{\mathrm{sp}}$.



We denote by $A_{s,t}$ the submatrix of $A$ associated with the cluster pair $(s, t)$. Once the block partitioning is established, admissible blocks $A_{s,t}$ are approximated using low-rank representations. In the $\mathcal{H}$-matrix format, each admissible block of size $m \times n$ and rank $k$ is represented as $A_{s,t} = U_{s,t} V_{s,t}^T$, where $U_{s,t}$ and $V_{s,t}$ are $m \times k$ and $n \times k$ matrices, respectively. This factorization leads to a storage complexity of $O(n \log n)$ for the entire matrix. The $\mathcal{H}^2$-matrix variant further reduces this complexity to $O(n)$ by employing a nested basis structure that enables shared representations across blocks.

Instead of storing independent $U$ and $V$ matrices for each admissible block, the $\mathcal{H}^2$-matrix format introduces a shared basis for block rows and columns associated with each cluster in the cluster tree. Each admissible block is then represented as $A_{s,t} = U_s S_{s,t} V_t^T$,
where $U_s$ and $V_t$ are the basis matrices associated with clusters $s$ and $t$, respectively, and a small ${k \times k}$ coupling matrix $S_{s,t}$. 

The bases for the leaf clusters are stored explicitly, while the basis for an internal node $\tau$ in the cluster tree is defined recursively in terms of the bases of its children $\tau_1$ and $\tau_2$ using transfer matrices. Specifically, the basis at $\tau$ is constructed via transfer matrices $E$ for $U$ and $T$ for $V$, resulting in a nested basis structure that enables efficient storage and computation:
\begin{equation*}
    \label{eqn:nested_basis}
    U_\tau = \begin{bmatrix}
    U_{\tau_1} & \\
     & U_{\tau_2}
    \end{bmatrix} \begin{bmatrix}
    E_{\tau_1} \\
    E_{\tau_2}
    \end{bmatrix} \quad \quad \quad
    V_\tau = \begin{bmatrix}
    V_{\tau_1} & \\
     & V_{\tau_2}
    \end{bmatrix} \begin{bmatrix}
    T_{\tau_1} \\
    T_{\tau_2}
    \end{bmatrix}
\end{equation*}

\begin{wrapfigure}[15]{r}{0.5\textwidth}
	\centering	\includegraphics[width=0.9\linewidth]{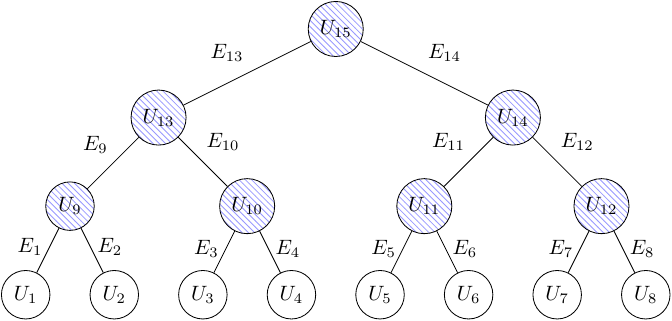}  
	\caption{The basis tree for the hierarchical matrix in Fig. \ref{fig:original_hmatrix_with_cluster_tree} where leaves $U_\tau$ are stored explicitly and the shaded inner nodes are implicitly represented by the nested bases using the transfer matrices $E$.}
	\label{fig:basis_tree}
\end{wrapfigure}

These notations are summarized in Table~\ref{tbl:notation}. We also use MATLAB notation where convenient. Figure~\ref{fig:basis_tree} illustrates the basis tree corresponding to the cluster hierarchy shown in Fig.~\ref{fig:original_hmatrix_with_cluster_tree}. In this representation, the clear nodes at the leaf level denote basis matrices that are stored explicitly, while the shaded internal nodes are defined implicitly via the nested basis property. For simplicity of exposition, we assume $A$ is symmetric and real-valued in the rest of this paper, leading to $V_{t}=U_{s}$ and $T_{t}=E_{s}$. However, the algorithm can be extended to nonsymmetric or complex-valued matrices, as in, 
for instance, \cite{Sushnikova2023}.

\begin{table}[b!]
\centering
\renewcommand{\arraystretch}{1.2} 
\begin{tabularx}{\textwidth}{@{} l X @{}}
\toprule
\textbf{Symbol} & \textbf{Description} \\
\midrule
$H$ & admissible part of a matrix \\
$H^l_{ij}$ & admissible block at level $l$ defined by clusters $i$ and $j$\\
$D$ & inadmissible part of a matrix \\
$F$ & working fill-in matrix\\
$U_i$, $V_i$       & bases of cluster $i$ in the hierarchical matrix \\
$E_i$, $T_i$       & transfer matrices for bases of cluster $i$ to its parent cluster \\
$S_{ts}$       & coupling matrix for the admissible block defined by clusters $t$ and $s$ \\
$A_\tau^l$       & matrix $A$ after skeletonization of cluster $\tau$ at level $l$ \\
$D_{i*}$  & all non-zero blocks in block row $i$ of a block sparse matrix $D$ \\
$D_{*j}$  & all non-zero blocks in block column $j$ of a block sparse matrix $D$ \\
$V^\bot$ & orthogonal complement of a matrix $V$ \\
$i_R$       & redundant indices of cluster $i$\\
$i_S$       & skeletonized indices of cluster $i$ \\
$\epsilon_{lu}$ & compression threshold for the fill-in \\
\bottomrule
\end{tabularx}
\caption{Definitions of symbols.}
\label{tbl:notation}
\end{table}

\subsection{Literature Review}

A number of works have addressed the direct factorization of \hh matrices with strong admissibility conditions, motivated by the needed to reduce rank growth that is encountered when weak admissibility structures are used, particularly in 3D problems.  A major line of work has been the use of skeletonization-based factorizations adapted to strong admissibility. 

Minden et al.~\cite{minden2017recursive}  introduced the strong recursive skeletonization (RS-S) factorization, generalizing the recursive skeletonization (RS) process of \cite{martinsson05}. RS-S 
 compresses only far-field interactions and leaves near-field interactions uncompressed and results in a hierarchical LU-like factorization expressed as a product of many block unit-triangular factors. It can achieve a slower rank growth than is possible with weak admissibility conditions, which compress all neighboring nodes. Under mild assumptions on numerical ranks, the complexity of RS-S scales linearly in the matrix size.

The strong-admissibility skeletonization idea has since been extended and applied in various contexts. Sushnikova et al.~\cite{Sushnikova2023} developed the FMM-LU solver which builds on the RS-S framework for 3D boundary integral equations. FMM-LU constructs an LU-like hierarchical factorization for dense matrices arising from elliptic and low-frequency wave equations on surfaces. By using a level-restricted octree and high-order quadrature, it was demonstrated that even complex 3D geometry problems can be factored in linear time with strong-admissibility compression, permitting efficient solution of large boundary-value problems. 
Yesypenko and Martinsson \cite{yesypenko2025} generalize RS-S via randomized sampling to make it blackbox and applicable to matrices available via matrix-vector products only. The method uses randomized SVDs, draws test vectors once up front, and injects structure dynamically as the factorization proceeds, yielding a sample‑efficient direct factorization even when individual matrix entries cannot be evaluated efficiently. 
Ma et al.~\cite{ma2022scalable,Ma_2024} also use RS-S but introduce a phase that pre-computes all possible fill-in contributions that may arise during the factorization and incorporates them into the basis matrices ahead of time, allowing the dependency on trailing submatrices to be removed. 
Ma and Jiao \cite{Ma2019} propose an RS-S variant that dynamically augments the original basis with new ones to account for the fill-in produced at every level, and a related version \cite{Ma_2024} that recomputes a new basis to account for the Schur complement updates to the original blocks during factorization.

Other approaches to direct factorization have also been proposed. Coulier at al.~\cite{coulier2017inverse} describe a strategy that uses a telescoping additive decomposition for applying the approximate inverse operator. The methods is termed an ``inverse fast multipole'' as it rewrites the dense system as an extended sparse system by introducing FMM multipole/local variables, then performs elimination while compressing and redirecting fill‑ins that occur between well‑separated clusters. B{\"o}rm and Reimer \cite{borm13_lr} propose methods based on matrix multiplication and local low rank updates that can be performed in linear complexity. These updates can be combined with a recursive procedure to approximate the product of two \hh matrices, and these products can be used to approximate the matrix inverse and the LU factorization. Although the asymptotic rates are favorable for these methods, their hidden constants are non-negligible in practice.











\subsection{Summary of Contributions}
We present a parallel fine-grained RS-S-style algorithm suitable for execution on GPUs and other massively parallel high-throughput architectures. Our algorithm is purely algebraic in its operations and runs as a black box starting only with an \hh matrix, facilitating its incorporation in libraries and computational workflows. The algorithm builds on the ideas in \cite{Ma2019}, and introduces a number of innovations to allow for scalable performance on modern hardware. It augments the \hh basis incrementally and adaptively to accommodate the necessary ``fill-in''. Our contributions may be summarized as follows: 

\begin{itemize}
	\item \emph{GPU-centric pipeline.} The computationally-demanding operations of the algorithm are all marshaled into batches and expressed in terms of batched linear algebra that can then be executed efficiently. In addition, dynamic memory allocations are completely avoided through the use of prefix-sum memory management operations that can also be efficiently executed on GPU architectures.
	
	\item \emph{Multi-level matrix graph coloring.} For every level of the matrix \wbnote{tree},
    a connectivity graph of its inadmissible blocks is partitioned into colors that allow for independent execution of the orthogonal projections and partial factorizations involved in the skeletonization  operations of matrix blocks at that level. This is an effective and scalable strategy as the number of colors does not grow with problem size.

	\item \emph{QR-based basis augmentation.} We propose and demonstrate the use of QR decompositions for building a basis for the fill-in induced by the Schur complement operations of the factorization. This is designed as a compromise between the potential loss of accuracy of Gram matrix approaches that square the condition number and the less-than-optimal ranks that may be produced by randomized SVDs. Increasingly on GPUs,  maximum performance is focused on single (and even lower) precision operations,  \deknote{putting a premium} on handling linear algebra operations with large condition numbers in a stable manner.
	
	\item \emph{Parallel solves.} In many settings the primary reason for a direct factorization is to allow for fast solves. This involves the application of the inverse factorization transformations in forward and backward substitution phases, which we cast in a computational structure similar to that of a hierarchical matrix-vector product, known to have effective parallel GPU execution \cite{boukaram19_gpu, h2opus}. As with the factorization, the solve phase involves splitting the application operations into smaller parallel conflict-free sub-batches whose effective execution is demonstrated. 
\end{itemize}

We present a battery of tests originating from 2D and 3D contexts to evaluate the performance of the algorithm. We are releasing the code and test cases as open-source for reproducibility and to address the practical need in the community for wider availability of readily usable source codes for factorizations of general \hh matrices with strong-admissibility structures.  

\section{Algorithms}

\subsection{Factorization Algorithm}

\begin{figure}[!htb] 
  \centering
  \includegraphics[width=\doublecolumnimgwidth]{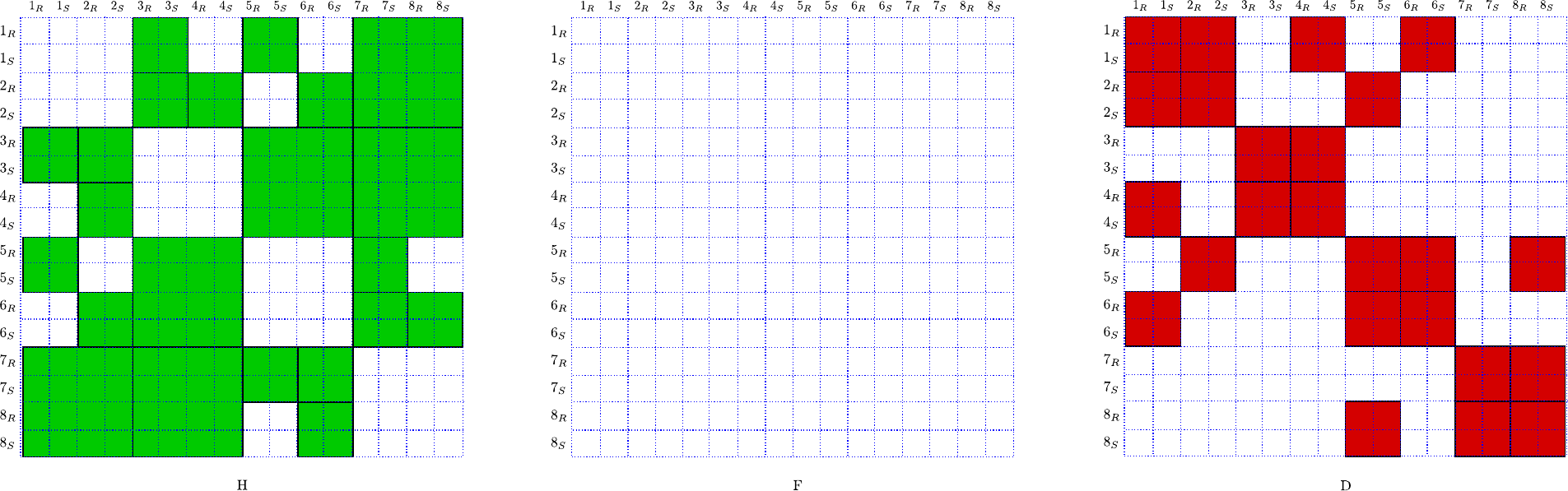}
  \caption{$A$ split into the admissible part $H$ in green, the inadmissible part $D$ in red and a fill-in matrix $F$ which is initially empty and whose blocks will be shown in black.
  }
  \label{fig:split_hmatrix}
\end{figure}
We use a skeletonization process
to generate approximations of blocks in the matrix by generating a basis that eliminates the redundant parts of the block while retaining representative skeletonized blocks. In sketching based approaches, this is typically achieved using interpolative decomposition to determine the redundant and skeleton indices. If we construct an orthogonal basis for our hierarchical matrix, we can use the complementary basis vectors to do the elimination instead. 

Let us split the hierarchical matrix $A$ in Fig. \ref{fig:original_hmatrix_with_cluster_tree} into three parts: the low rank part $H$, the dense part $D$ and a fill-in matrix $F$ which is initially the zero matrix: $A = H + D + F$ as shown in Fig. \ref{fig:split_hmatrix}. Any transformation applied to $A$ will be applied to all three of its parts and after each transformation, we are free to shuffle around the blocks between the three parts as needed (for example moving the sum of certain blocks from $F$ and $H$ to $D$). Let $m$ be the leaf size and $k^l$ be a \wbnote{representative rank of the blocks at level $l$.} 

Factorization using skeletonization
follows two major phases: orthogonal projections to eliminate the redundant rows and columns of a cluster within $F$ and $H$ and partial factorization in $D$ to eliminate the same rows and columns without allowing additional fill-in from previous steps to affect $F$ and $H$.  Let us call the transformed matrix at level $l$ after cluster $\tau$ has been skeletonized $A_\tau^l$. The first phase will often involve computing a basis that will include the original cluster basis as well as a basis that approximates the blocks in $F$. Starting from the leaf level, cluster 1 of the matrix is skeletonized as follows:



\begin{figure}[!htb]
  \centering
  \includegraphics[width=\doublecolumnimgwidth]{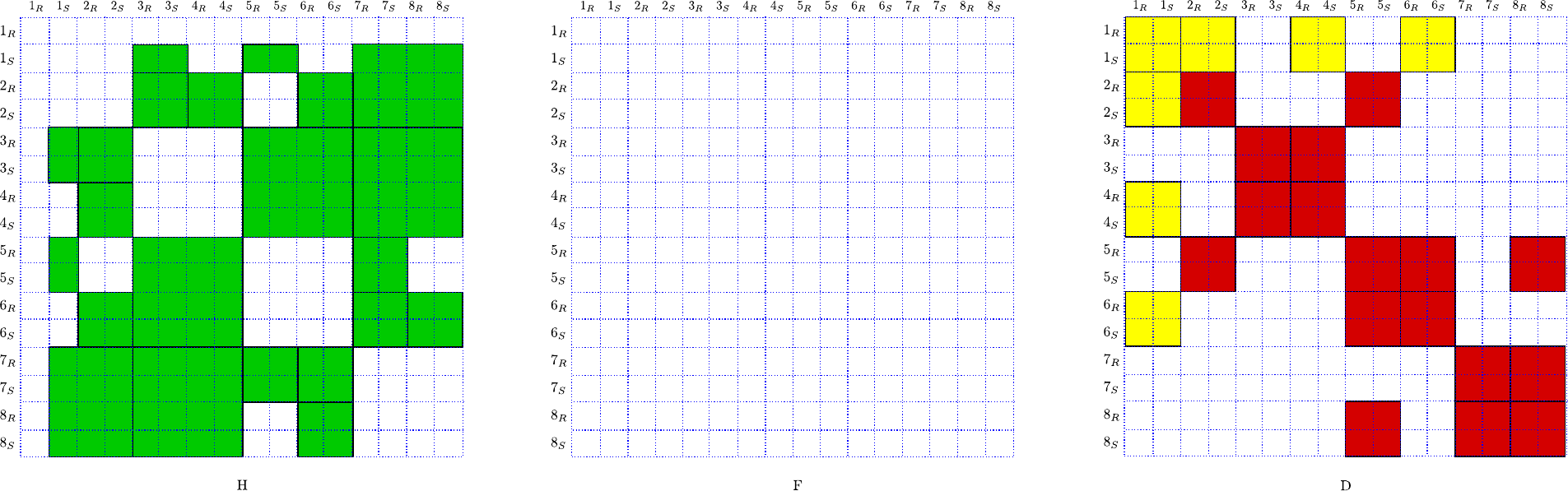}
  \caption{The effect of the orthogonal projection generated from cluster 1 on the three parts of $A$, where the rows and columns $1_R$ of $H$ are zeroed out and the first block row and column of $D$, shown in yellow, are scaled by $\tilde{Q}_1$.}
  \label{fig:cluster_01_step_02}
\end{figure}

\begin{figure}[!htb]
  \centering
  \includegraphics[width=\doublecolumnimgwidth]{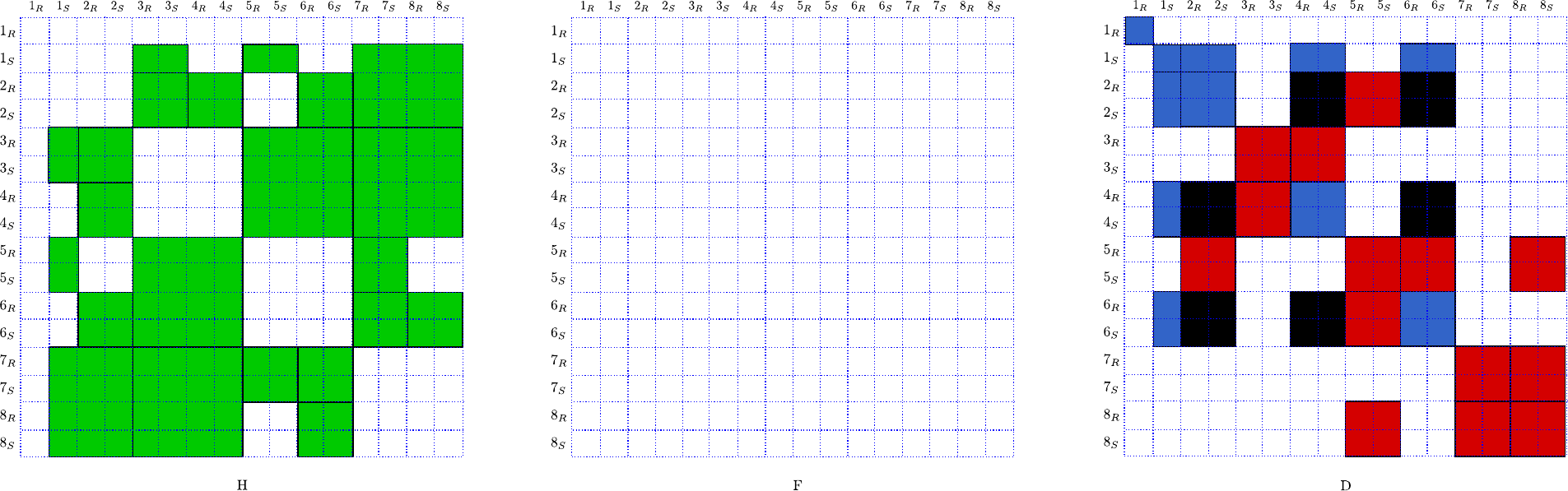}
  \caption{Partial factorization of $D$ to eliminate the off diagonal blocks in the $1_R$ row and columns. The fill-in due to the updates are shown in black and updated blocks of the original matrix are in blue.}
  \label{fig:cluster_01_step_03}
\end{figure}

\begin{enumerate}
    \item We first complete the cluster basis $V_1$ with its complement $V_1^\bot$ to form the matrix $\tilde{Q}_1 = [V_1^\bot, V_1]$. 
    \item The block row and column corresponding to the first cluster are then scaled by $\tilde{Q}_1$ by forming the product $A_1^1 = Q_1^T A Q_1 = Q_1^T (H+D+F) Q_1$, where $Q_1 = \texttt{diag}(\tilde{Q}_1, I, \cdots, I)$. This zeros out the rows and columns of $H$ corresponding to the redundant portion of the cluster which we call $1_R$, since $\tilde{Q}_1^T V_1 = [0; I]$. The fill-in matrix is still the zero matrix at this point. 
    The block row and column $D_{1*}$ and $D_{*1}$, highlighted in yellow in Fig. \ref{fig:cluster_01_step_02},
    are scaled by $\tilde{Q}^T_1$ and $\tilde{Q}_1$ respectively, 
    i.e., $\tilde{Q}_1^T D_1^1 \tilde{Q}_1$. 
    \item The off-diagonal blocks in the block row and column $1_R$ of $D$ are eliminated using partial LU factorization. The diagonal block $D_{1_R 1_R}$ is factorized using dense LU decomposition and elementary lower and upper block triangular matrices are constructed to eliminate the redundant off-diagonal blocks $D_{1_R *}$ and $D_{*1_R}$ :
    \begin{equation*}
    L_1 = \begin{bmatrix}
    I & & & \\
    -D_{1_S1_R}D^{-1}_{1_R1_R} & I & & \\
    -D_{*1_R} D^{-1}_{1_R1_R} & & I
    \end{bmatrix} 
    \quad \quad
    U_1 = \begin{bmatrix}
    I & -D^{-1}_{1_R1_R}D_{1_R1_S} & -D^{-1}_{1_R1_R}D_{1_R *}& \\
     & I & & \\
     & & I
    \end{bmatrix} 
    \end{equation*}
    The matrix can then be transformed as \wbnote{$A^1_1 \rightarrow L_1 A^1_1 U_1$,} 
    leaving only the diagonal block $D_{1_R1_R}$ in the redundant part of the inadmissible matrix. Since the redundant block rows and columns are zero in $H$ and $F$, they are unaffected by the row and column operations performed during this step, whereas fill-in is introduced in $D$ due to the Schur complement updates. \ylnote{By construction, the fill-in blocks are always admissible}. This is shown in Fig. \ref{fig:cluster_01_step_03} with fill-in in black and updated original blocks are in blue. 
    \item The fill-in introduced in the previous step is moved from $D$ to $F$. \wbnote{Though they are formally dense, they are low rank by construction as the outer product of the cluster's redundant off-diagonal blocks}. If we can ensure that the redundant block rows and columns in $F$ are eliminated at the same time as the elimination in $H$ during the projection phase, this will prevent the fill-in produced by the partial factorizations from creating more fill-in as we proceed to other clusters. Figure \ref{fig:cluster_01_step_03_fillin} shows all three parts of the updated matrix at the end of processing cluster 1.
\end{enumerate}

Moving on to the second cluster, we must determine additional orthogonal basis vectors that allow us to eliminate the redundant portions of the fill-in blocks in the rows and columns corresponding to that cluster. Consider the fill-in block row $F_{2*} = [F_{24}, F_{26}]$ in Figure \ref{fig:cluster_01_step_03_fillin}. We would like to find an orthogonal basis $\tilde{V}_2 = [V_2, \bar{V}_2]$ 
that includes the original basis $V_2$ and produces an approximation of $F_{2*} \approx \tilde{V}_2 \tilde{V}^T_2 F_{2*}$, minimizing the error $||\tilde{V}_2 \tilde{V}^T_2 F_{2*} - F_{2*}||$. This can be achieved by computing the SVD of the matrix $Y = (I-V_2V_2^T)F_{2*} = \bar{V}_2 S W_2^T$, since the computed singular vectors $\bar{V}_2$ minimize the error:

\begin{figure}[!htb]
  \centering
  \includegraphics[width=\doublecolumnimgwidth]{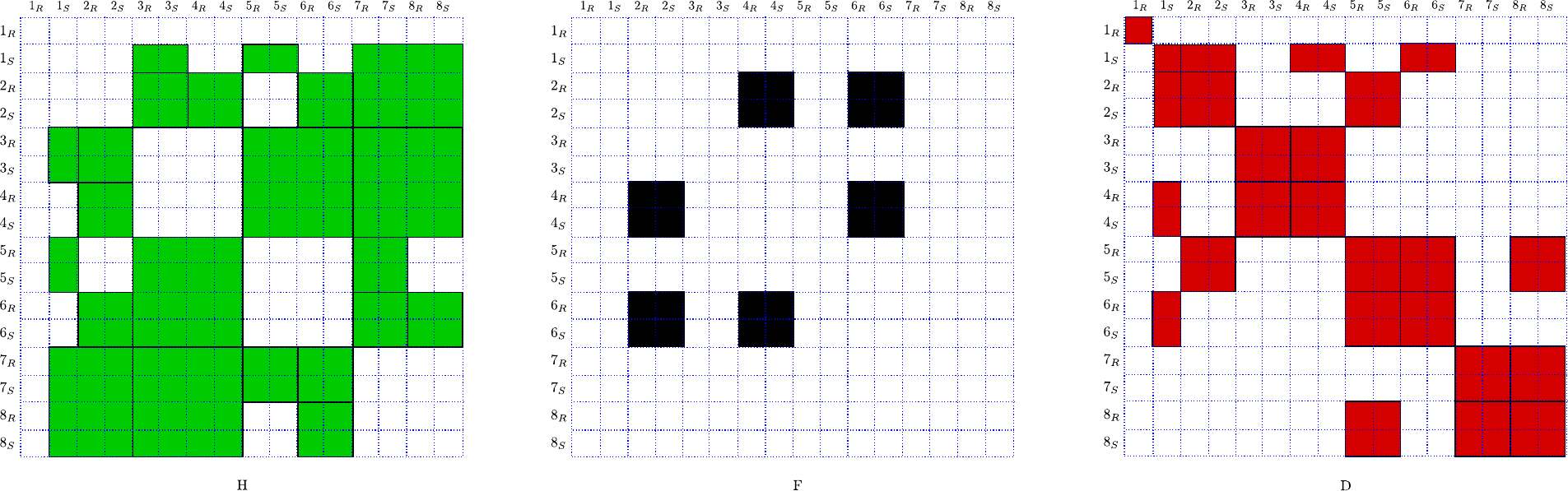}  
  \caption{Moving the black fill-in blocks generated by the partial factorization of $D$ to the fill-in matrix $F$.}
  \label{fig:cluster_01_step_03_fillin}
\end{figure}



\begin{figure}[!htb]
  \centering
  \includegraphics[width=\doublecolumnimgwidth]{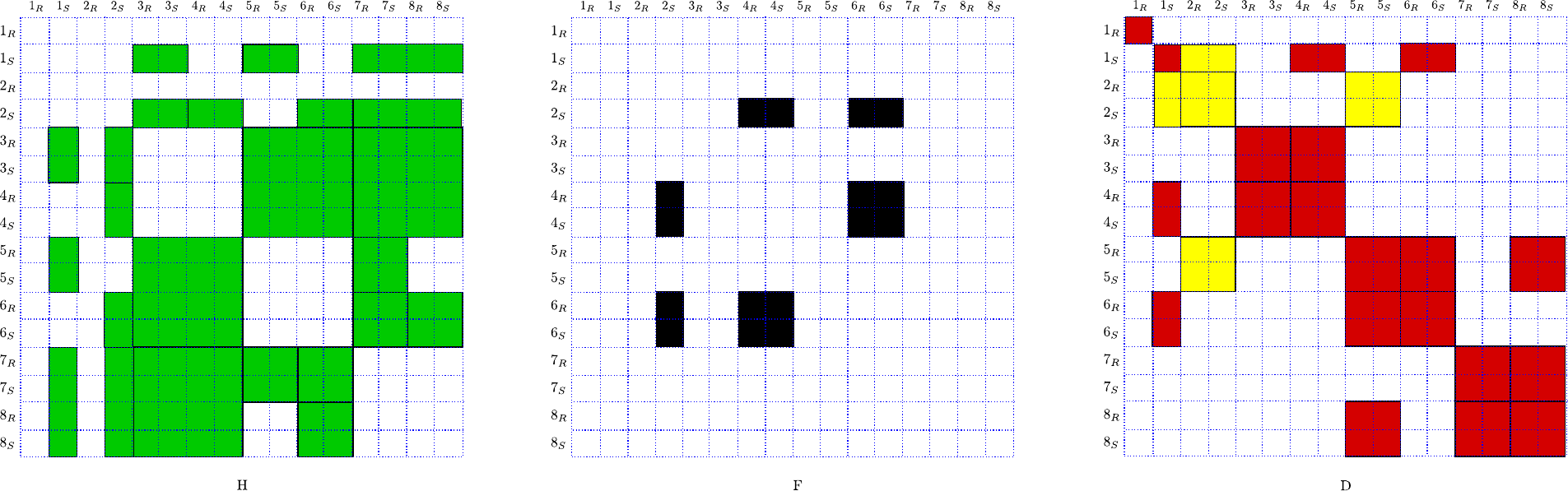}  
  \caption{Skeletonizing the second cluster. The projection matrix for this step includes additional basis vectors for the fill-in blocks in block row 2 of $F$. This allows the rows and column $2_R$ in both $H$ and $F$ to be skeletonized at the same time.}
  \label{fig:cluster_02_step_02}
\end{figure}

\begin{figure}[!htb]
  \centering
  \includegraphics[width=\doublecolumnimgwidth]{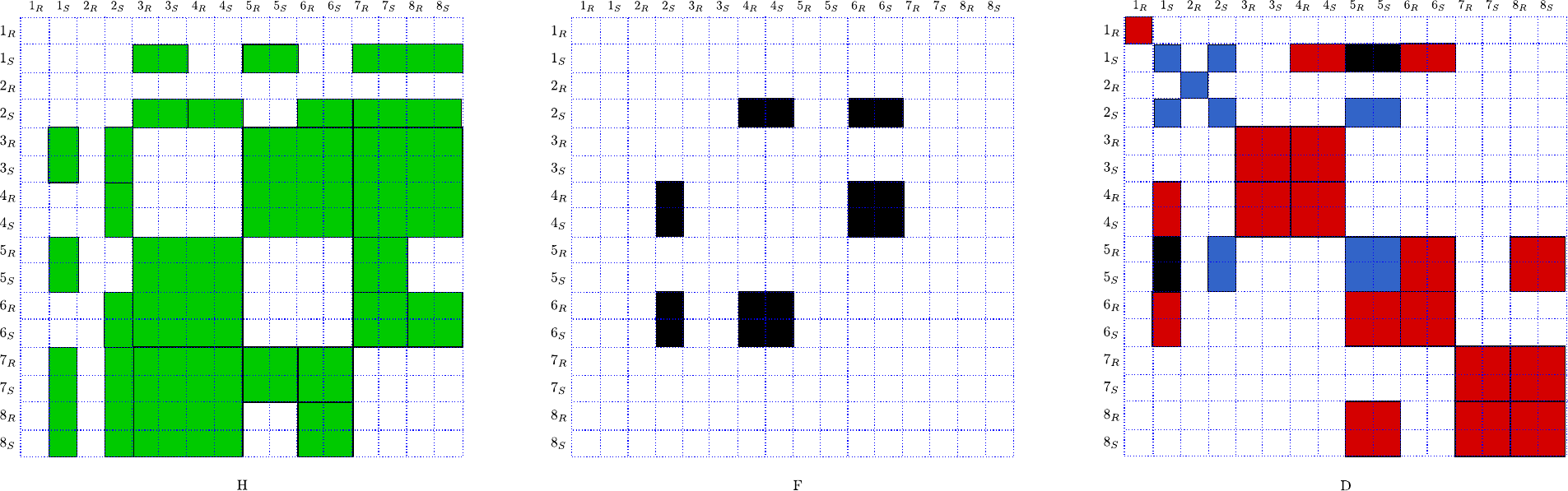}  
  \caption{The partial factorization of the second cluster in $D$ generates rectangular fill-in when eliminating the blocks $D_{2_R 1_S}$ and $D_{1_S 2_R}$ .}
  \label{fig:cluster_02_step_03}
\end{figure}

\begin{equation*}
\begin{split}
||\bar{V}_2 \bar{V}_2^T (I-V_2V_2^T)F_{2*} - (I-V_2V_2^T)F_{2*}|| & = ||\bar{V}_2 \bar{V}_2^T F_{2*} - \bar{V}_2 \cancelto{0}{\bar{V}_2^T V_2} V_2^T F_{2*} - F_{2*} + V_2V_2^T F_{2*}|| \\ 
& = || [V_2, \bar{V}_2] [V_2, \bar{V}_2]^T F_{2*} - F_{2*} || 
 = ||\tilde{V}_2 \tilde{V}^T_2 F_{2*} - F_{2*}||
\end{split}
\end{equation*}

We then truncate \deknote{the singular vectors of $\bar{V}_2$ that are unnecessary to satisfy a} compression threshold $\epsilon_{fill} = \epsilon_{lu} ||A||$. %
As for computing the SVD of $Y$, there are at least three algorithm choices:
1) one can form the Gram matrix $G = YY^T$ and find its SVD and truncate to $\epsilon_{fill}^2$;
2) perform the QR decomposition of $F_{2*}^T = QR$ and then the SVD of $R$; or
3) use a randomized SVD. 
In our tests, we use the second approach since it will provide the best accuracy, but the Gram matrix approach can trade accuracy for speed, while the randomized SVD can lead to slightly larger ranks but is more stable, providing a middle ground between the two. Regardless of the choice of compression algorithm, we end up with additional basis vectors $\bar{V}_2$ and an augmented cluster basis $\tilde{V}_2 = [V_2, \bar{V}_2]$. To account for the additional basis vectors, we simply need to pad the original coupling matrices with zeros, increasing the rank $k^1$: 
\begin{equation*}
  V_i S_{ij} V_j^T = [V_i, \bar{V}_i] \begin{bmatrix} S_{ij} & 0 \\ 0 & 0 \end{bmatrix} [V_j, \bar{V}_j]^T
  = \tilde{V}_i \tilde{S}_{ij} \tilde{V}_j^T
\end{equation*}

To maintain the nested basis property, the rows of the corresponding transfer matrices $T_i$
are padded with zeros as well, transforming into $\tilde{T}_i = [T_i; 0]$.

The fill-in in $F$ for the second cluster can now be skeletonized at the same time as the second block row and columns in $H$. We again form the projection matrix by finding the complement $\tilde{V}_2^\bot$ to form the matrix $\tilde{Q}_2 = [\tilde{V}_2^\bot, \tilde{V}_2]$. As with the first cluster, scaling the matrix as $A_2^1 = Q_2^T A_1^1 Q_2$ where $Q_2 = \texttt{diag}(I, \tilde{Q}_2, \cdots, I)$ eliminates the redundant portion $2_R$ of the cluster rows and columns since $\tilde{Q}_2^T \tilde{V}_2 = [0; I]$. It also eliminates the same rows and columns of the fill-in matrix, since we approximate each block in that cluster as $F_{2j} \approx \tilde{V}_2 \tilde{V}^T_2 F_{2j}$:

\begin{equation*}
\tilde{Q}_2^T F_{2j}  \approx [\tilde{V}_2^\bot, \tilde{V}_2]^T \tilde{V}_2 \tilde{V}^T_2 F_{2j}
    = \begin{bmatrix} 0 \\ I \end{bmatrix} \tilde{V}^T_2 F_{2j} = \begin{bmatrix} 0 \\ \tilde{V}^T_2 F_{2j} \end{bmatrix}
\end{equation*}

Figure \ref{fig:cluster_02_step_02} shows the skeletonization of $H$ and $F$ as well as the scaled blocks of $D$ in yellow. The partial LU factorization then proceeds as before with fill-in generated in $D$ as shown in 
Fig.~\ref{fig:cluster_02_step_03}. Elementary triangular matrices $L_2$ and $U_2$ are computed and used to transform the matrix as \wbnote{$A_2^1 \rightarrow L_2A_2^1U_2$}, 
as with the first cluster. Note that rectangular fill-in can be produced due to the elimination of previously skeletonized clusters and can simply be moved to $F$ after the updates are complete.

\begin{figure}
  \centering
  \includegraphics[width=\doublecolumnimgwidth]{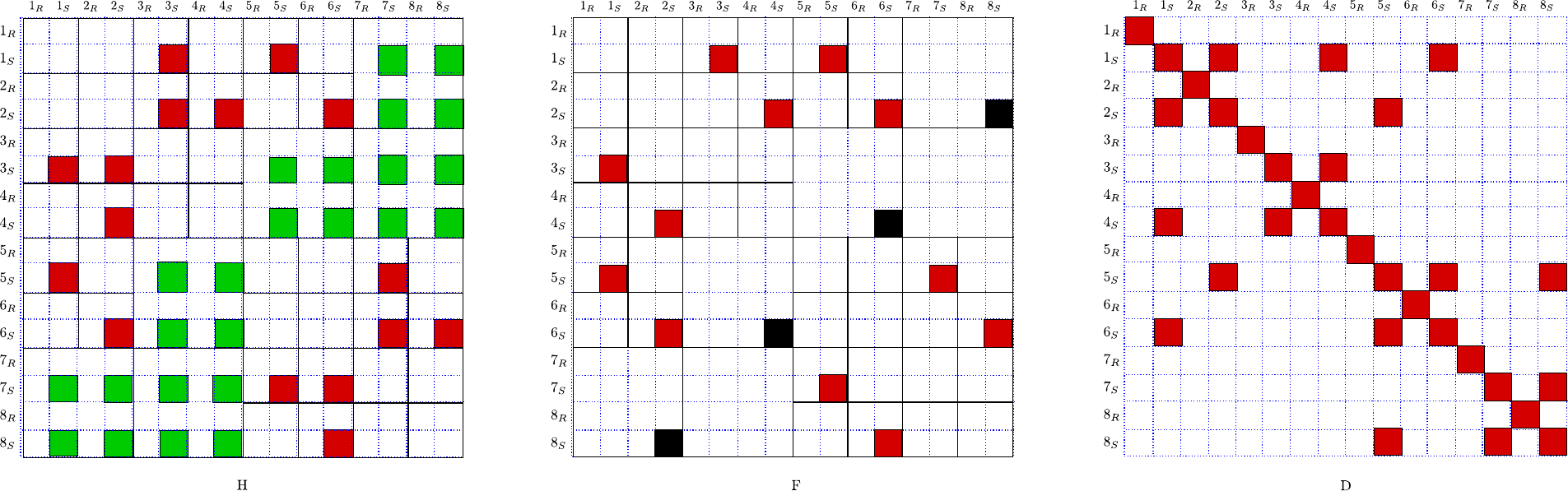}  
  \caption{All clusters have been skeletonized and updated. The blocks at the same level of $H$ and $F$ can now be treated as dense blocks and are colored red.}
  \label{fig:level_02_expand}
\end{figure}
\begin{figure}
  \centering
  \includegraphics[width=\doublecolumnimgwidth]{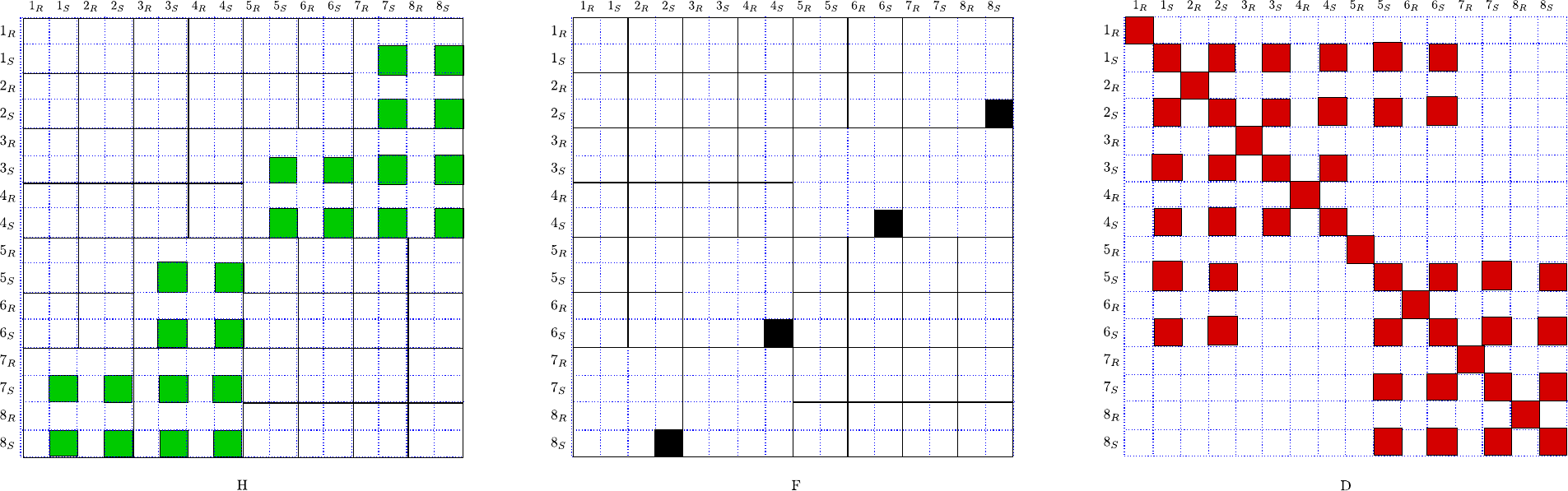}  
  \caption{Adding the converted blocks from $H$ and $F$ and moving them to $D$.}
  \label{fig:level_02_exchange}
\end{figure}
\begin{figure}
  \centering
  \includegraphics[width=\doublecolumnimgwidth]{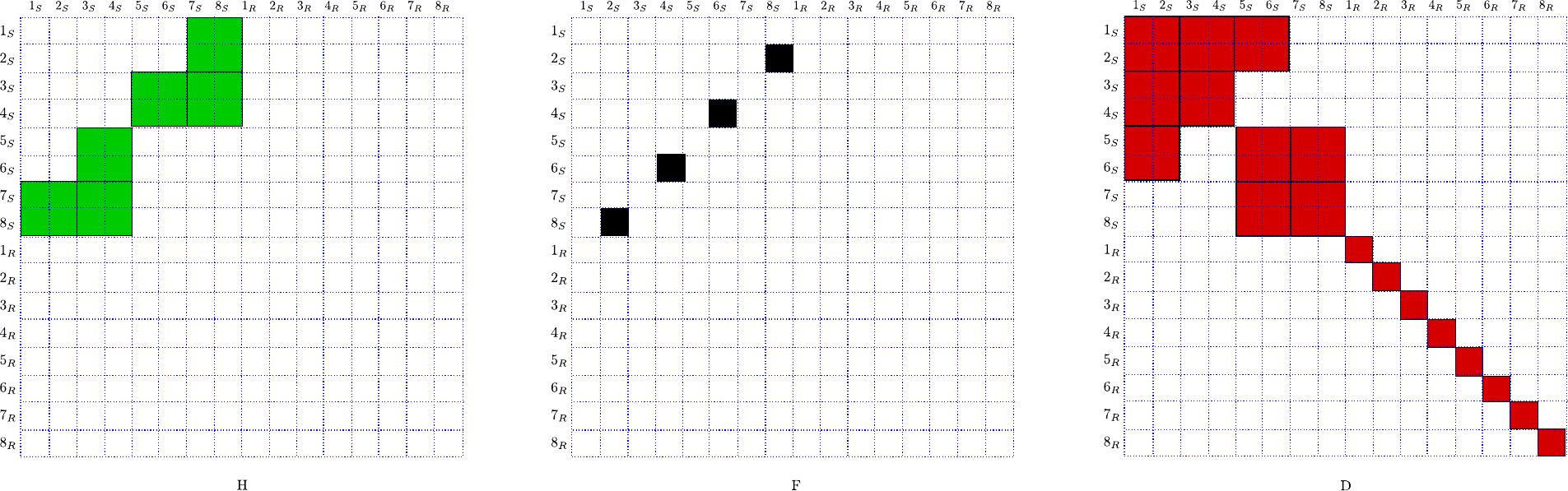}  
  \caption{Permuting all three matrices to split the $S$ and $R$ indices. The resulting matrix is a $2 \times 2$ block diagonal. The lower right block is also block diagonal, so inverting it is trivial. The upper right block can then be processed using the same method recursively.}
  \label{fig:level_02_permute}
\end{figure}

Once all eight clusters have been skeletonized, we end up with the matrix $A_8^1$ as the sum of the three matrices in Figure \ref{fig:level_02_expand}. At this point, \wbnote{all blocks $H^1_{ij}$ at the leaf level of $H$} have been transformed as:
\begin{equation*}
\begin{split}
  \tilde{Q}^T_i H^1_{ij} \tilde{Q}_j = \begin{bmatrix}\tilde{V}^\bot_i & \tilde{V}_i\end{bmatrix}^T \tilde{V}_i \tilde{S}_{ij} \tilde{V}_j^T \begin{bmatrix}\tilde{V}^\bot_j & \tilde{V}_j\end{bmatrix} 
  &= \begin{bmatrix}0_{(m-k^l)\times k^l} \\ I_{k^l \times k^l} \end{bmatrix}\tilde{S}_{ij} \begin{bmatrix}0_{k^l\times(m-k^l)} & I_{k^l \times k^l}\end{bmatrix} \\
  &= \begin{bmatrix}0_{(m-k^l)\times (m-k^l)} & 0_{(m-k^l)\times k^l} \\ 0_{k^l\times (m-k^l)} & \tilde{S}_{ij} \end{bmatrix} 
\end{split}
\end{equation*} 
Now each admissible block $H^1_{ij}$ can be expanded as a dense block containing only $k^l \times k^l$ non-zeros in the skeletonized rows and columns of the matrix $i_S$ and $j_S$. The same applies for the fill-in matrix $F$. 
Since we can freely move blocks between the three matrices, we select all blocks $H^1_{ij}$ from $H$ and the corresponding blocks $F_{ij}$ from $F$ as shown in red in Fig. \ref{fig:level_02_expand}, add them together and move them to $D$ as shown in Fig. \ref{fig:level_02_exchange}. 
Since we have removed all blocks $H^1_{ij}$ from $H$ and the basis clusters $U_i$ and $U_j$ have all taken the form $[0; I]$, we can completely remove the first level of the basis from $H$. Let ${\tau_1}$ and ${\tau_2}$ be the children of a node ${\tau}$ at the next level of the matrix. The basis node $V_\tau$, originally expressed in terms its children $V_{\tau_1}$ and $V_{\tau_2}$ and two transfer matrices $T_{\tau_1}$ and $T_{\tau_2}$, is transformed as:
\begin{equation*}
    \begin{bmatrix} \tilde{Q}_{\tau_1}^T \\ \tilde{Q}_{\tau_2}^T\end{bmatrix} V_\tau = 
    \begin{bmatrix}
    \tilde{V}^\bot_{\tau_1} & \tilde{V}_{\tau_1} \\ 
    \tilde{V}^\bot_{\tau_2} & \tilde{V}_{\tau_2}\end{bmatrix}^T \begin{bmatrix}\tilde{V}_{\tau_1} & 0 \\ 0 & \tilde{V}_{\tau_2}\end{bmatrix}\begin{bmatrix} \tilde{T}_{{\tau_1}} \\ \tilde{T}_{{\tau_2}}\end{bmatrix} = \begin{bmatrix} \begin{bmatrix} 0 \\ I \end{bmatrix} \tilde{T}_{{\tau_1}} \\  \begin{bmatrix} 0 \\ I \end{bmatrix} \tilde{T}_{{\tau_2}}\end{bmatrix} = \begin{bmatrix} 0 \\ \tilde{T}_{{\tau_1}} \\ 0 \\ \tilde{T}_{{\tau_2}}\end{bmatrix} 
\end{equation*}
The zero blocks correspond to the redundant indices in the matrix. When we separate the redundant indices from the skeletonized ones by applying a permutation $P^l$ at level $l$, the resulting matrices are shown in Fig. \ref{fig:level_02_permute}. All matrices are now $2 \times 2$ block diagonal matrices, with the lower right blocks of $H$ and $F$ as zero blocks. $H$ maintains its properties as a hierarchical matrix, with the basis at its leaf level now consisting only of the transfer matrices
$    V_t = \begin{bmatrix} \tilde{T}_{{\tau_1}} \\ \tilde{T}_{{\tau_2}}\end{bmatrix} $
and its coupling matrices untouched. Once all clusters in a level have been skeletonized, we end up with the matrix $A_8^1 = \left(L_8Q^T_8 \dots L_1Q_1^T \right)A \left(Q_1 U_1 \dots Q_8 U_1\right)$.
Note that each factor in the transformation is readily invertible, either as the transpose of the orthogonal matrices $Q$ or by flipping the sign of the off-diagonal blocks in the elementary triangular matrices. The lower right block of $D$ is block diagonal, so we can invert it by applying a block diagonal factorization of the redundant blocks as $L_r^lU_r^l$, whereas the upper left blocks can then be processed in the next level. The computation continues level by level until we reach the highest level $t$ of the tree that contained admissible blocks. At that point, only inadmissible blocks remain and the matrix $D_t$ is a relatively small dense matrix that can be factorized directly into its factors $D_t=L_tU_t$, ending the factorization. The final factorization is therefore
\begin{equation}
A = \mathcal{L}\mathcal{U} = \left( \prod_{l=1}^t \left(\prod_{\tau \in I^l} Q_\tau L_\tau \right) L_r^l P^{l^T} \right) L_t U_t \left( \prod_{l=1}^t \left( P^l U_r^l \prod_{\tau \in I^l}U_\tau Q_\tau^T \right) \right).
\label{eq:factorization}
\end{equation}
Algorithm \ref{alg:skelfactor} describes the entire factorization algorithm starting at the leaf level up to the top level where the final dense factorization takes place, while Algorithm \ref{alg:skelcol} describes the parallel skeletonization of all clusters 
which we describe in the following section.

\begin{algorithm}
\caption{\textsc{factorize}($A$, $\epsilon_{lu}$): Factorize a hierarchical matrix using skeletonization}
\label{alg:skelfactor}
\begin{algorithmic}[1]
  \Require $\mathcal{H}^2$-matrix $A$, truncation threshold $\epsilon_{lu}$
  \State $d = \texttt{depth}(A)$
  \State $t = \texttt{topLevel}(A)$ \Comment{First level that has admissible nodes}
  \State $F = 0$  \Comment{Initially zero fill-in matrix}
  \State $Z = []$ \Comment{Empty factorization}
  \For{$l \gets d $ to $t$}
    \State $D = \texttt{inadmissible}\left(A, F, l \right)$ \Comment{Extract the inadmissible nodes at level $l$}  
    \State $F = \texttt{fillIn}\left(A, F, l \right)$ \Comment{Extract the fill-in to higher levels} 
    \State $C^l = \texttt{coloring}\left(A, l \right)$  \Comment{Color nodes to determine independent skeletonization indices}
    \For{$C \in C^l$}
        \State $D, F, L, U, V_D = $ \texttt{skeletonizeColor}($A$, $D$, $F$, $C$, $\epsilon_{lu}$)
        \State $Z = \texttt{addFactor}(Z, C, L, U, V_D)$
    \EndFor
  \EndFor
  \State $D_t = \texttt{inadmissible}\left(A, F, t \right)$ \Comment{Extract the dense top level remnants}
  \State $L_t, U_t = \texttt{LU}(D_t)$
  \State $Z = \texttt{addTopLevel}(Z, L_t, U_t)$
  \State \Return $Z$
\end{algorithmic}
\end{algorithm}

\begin{algorithm}[b]
\caption{\textsc{skeletonizeColor}($A$, $D$, $F$, $C$, $\epsilon_{lu}$): Skeletonize the colored nodes}
\label{alg:skelcol}
\begin{algorithmic}[1]
  \Require $\mathcal{H}^2$-matrix $A$, inadmissible blocks $D$, fill-in matrix $F$
  \Require Color node indices $C$, truncation threshold $\epsilon_{lu}$
  \State $Q = I$  
  \ParallelFor{$i \in C$}
    \State $\bar{V}_i = \texttt{truncSVD}\left((I - V_iV_i^T) F_{i*}, \epsilon_{lu} \right)$ \Comment{Compute truncated left singular vectors}  
    \State $\tilde{V}_i = \begin{bmatrix} V_i & \bar{V}_i \end{bmatrix}$  \Comment{Augment the basis to accommodate fill-in}
    \State $\tilde{V}^\bot_i = $ \texttt{complement}$\left( \tilde{V}_i \right)$     \Comment{Get the vectors orthogonal to the augmented basis}
    \State $\tilde{Q}_i = \begin{bmatrix} \tilde{V}^\bot_i & \tilde{V}_i \end{bmatrix}$   \Comment{Complete the basis}  
    \State $Q\left(i, i \right) = \tilde{Q}_i$  \Comment{Set the diagonal block for node $i$}
  \EndParallelFor
  \State $D = Q^T D Q$
  \State $F = Q^T F Q$
  \State $\left(D, L, U, F^a \right) = \texttt{partialLU}\left( D, C \right)$ \Comment{Generate fill-in and elementary triangular matrices}
  \State $F = F + F^a$
  \State \Return $\left(D, F, L, U, Q \right)$
\end{algorithmic}
\end{algorithm}

\subsection{Parallelization}
An important property of the factorization algorithm is that, since we skeletonize the fill-in at the same time as the admissible matrix, fill-in due to the partial elimination of one cluster will affect only the rows and columns of the cluster's direct non-zero neighbors in the block sparse matrix. Any two clusters that are not directly connected to each other by a non-zero block can therefore be skeletonized independently. 
Forming the connectivity graph of the inadmissible block sparse matrix at each level, we can therefore apply a graph coloring algorithm to determine partitions of clusters that can be skeletonized in parallel. 
While each color can be processed in parallel, the number of colors will dictate the number of serial steps for each level of the tree. Since the maximum number of colors is determined by the maximum degree of the connectivity graph and the degree is the sparsity constant, the number of serial steps does not grow with the problem size. This should allow the algorithm to scale well with the problem size.

While the current results are obtained on a CPU in a development phase, we designed the algorithm with efficient GPU execution (and fine-grained parallelism) in mind. Since all of the blocks involved in the computation are quite small, either as the leaf size of the cluster or as the rank of the blocks, we first marshal the operations for all the clusters of the same color into batches that can be executed efficiently using batched linear algebra routines. 
Memory allocations are also kept to a minimum by computing matrix memory requirements for all clusters involved in a color's skeletonization, allocating a single large buffer and then using prefix sums on the products of the block dimensions to determine offsets into the buffer. Since workspace requirements are not known beforehand, we avoid costly allocations and de-allocations by using a custom memory pool allocator for all temporary allocations.

Though some operations are available on the GPU in a batched form, the GPU implementation of the missing operations (such as a non-uniform batched SVD for basis augmentation) remain as future work. On the CPU these batched operations are simple parallel OpenMP loops around single-threaded linear algebra kernels, while the marshaling is handled by Thrust routines with the OpenMP execution backend. 
A complication that must be addressed is that although the scaling of the rows and columns by the completed orthogonal basis is trivial, the partial factorization potentially involves multiple clusters of the same color trying to update the same block at the same time. To overcome this, we split the batched update into multiple smaller sub-batches that can be completed in parallel without any race conditions. The sub-batches are then executed in serial order. It is worth noting that since many of the operations are potentially quite small (for example, due to low ranks), they may be limited by the available memory bandwidth rather than by compute power.

\subsection{Forward and Backward Solves}
As noted in the previous section, all transformations applied to the matrix $A$ in Eq.~(\ref{eq:factorization}) to skeletonize and factorize it are easily invertible. To solve the system of equations represented by $A$ with a right hand side $b$, we simply need to apply the inverse of those transformations in the forward and backward substitution phases akin to the regular dense case, but with a computational structure similar to that of a hierarchical matrix-vector product. 
In the forward solve, we compute a vector tree $\hat{b}$ that shares the same structure as our basis tree. One can think of each level $\hat{b}^l$ as a vector of the same dimension as the skeletonized matrix $A^l$, with the leaf level of the tree initialized with our input vector $b$. The vector $\hat{b}^l$ is scaled by the orthogonal factor $Q_\tau$ followed by the elementary factor $L_\tau^{-1}$ which is simply $L_\tau$ with the sign of the off-diagonal blocks flipped. After all level transformations are completed, the block diagonal solves are carried out and the skeletonized portions can be swept up to the next level of $\hat{b}$.
At the top level $t$ a regular dense forward solve with $L_t$ and the tree vector $\hat{b}^t$ ends the forward solve phase. This is followed by a regular backwards solve with $U_t$, leading us to the start of the downwards sweep of the tree in the backwards solve phase. This is similar to the previous phase, starting at level $t$, applying the inverse of the transformations and sweeping down the tree at the end of the level. Once we reach the leaf level, we have our final solution vector $x$ as the leaf level of the downswept $\hat{b}$. The hierarchical solve routine is described in Algorithm \ref{alg:skelsolve}. The upsweep and downsweep at each level encapsulate the permutation process shown in Fig.~\ref{fig:level_02_permute} and its inverse, respectively, while the diagonal solve steps apply the inverses of the lower and upper diagonal redundant blocks. Finally, the \texttt{applyForward} and \texttt{applyBackward} routines apply the orthogonal factors and the elementary triangular factors for all the nodes of a specified color in parallel. Like the factorization, this operation involves splitting the application of the inverse of the elementary triangular operations into smaller parallel conflict-free sub-batches that are executed one after the other. 

\begin{algorithm}
\caption{\textsc{solve}($Z$, $b$): Solve $Ax=b$ using the factorization $Z$ of $A$}
\label{alg:skelsolve}
\begin{algorithmic}[1]
  \State $d = \texttt{depth}(Z)$
  \State $t = \texttt{topLevel}(Z)$ \Comment{First level that has admissible nodes}
  \State $\hat{b}^d = b$            \Comment{Initialize leaf level with RHS}
  \For{$l \gets d $ to $t+1$}         \Comment{Forward upsweep solve}
    \State $C^l = \texttt{coloring}\left(Z, l \right)$  \Comment{Color nodes}
    \For{$C \in C^l$}
        \State $\hat{b}^l = \texttt{applyForward}(Z, C, \hat{b}^l)$
    \EndFor
    \State $\hat{b}^l = \texttt{diagonalForward}(Z, \hat{b}^l)$   \Comment{Forward solve with redundant diagonal blocks}
    \State $\hat{b}^{l-1} = \texttt{upsweep}(Z, \hat{b}^l)$     \Comment{Sweep skeletonized vector to parent level}
  \EndFor
  \State $\hat{b}^t = \texttt{solve}(Z.L_t, Z.U_t, \hat{b}^t)$ \Comment{Dense top level solve}
  
  \For{$l \gets t+1 $ to $d$}         \Comment{Backwards downsweep solve}
    \State $\hat{b}^l= \texttt{downsweep}(Z, \hat{b}^{l-1})$     \Comment{Sweep skeletonized vector to child level}
    \State $\hat{b}^l = \texttt{diagonalBackward}(Z, \hat{b}^l)$   \Comment{Backward solve with redundant diagonal blocks}
    \State $C^l = \texttt{coloring}\left(Z, l \right)$  \Comment{Color nodes}
    \For{$C \in C^l$}
        \State $\hat{b}^l = \texttt{applyBackward}(Z, C, \hat{b}^l)$
    \EndFor
  \EndFor
  \State \Return $\hat{b}^d$ \Comment{The solution is the leaf level}
\end{algorithmic}
\end{algorithm}

\subsection{Algorithmic Complexity}
We consider the sparsity constant $C_{sp}$ for the inadmissible block sparse matrices at each level of the matrix tree and a representative rank $k$ for the basis of the clusters. At the leaf level, the cluster size is equal to the leaf size which we will assume to be a constant on the same order as $k$ and at higher levels they are $2k \times 2k$ blocks, so we will assume that all blocks are of size $O(k)$. 
The computational effort for the skeletonization of a cluster $\tau$ can be split into three major steps: finding the augmented basis, applying the complete orthogonal factor to the inadmissible and fill-in matrices and the partial LU factorization. The first step involves the compression of the $C_{sp}$ fill-in blocks in the cluster's block row $F_{\tau *}$ at a cost $O(C_{sp}k^3)$ with a similar cost for applying the orthogonal projection to the matrix blocks. 
On the other hand, the partial LU factorization involves $O(C_{sp}^2)$ blocks during the Schur complement updates, dominating the cost at $O(C_{sp}^2k^3)$ operations. Since there are $O(n)$ clusters in total, the cost of the factorization algorithm is therefore $O(C_{sp}^2k^3n)$ operations. 
The solve phase is similarly bound by the cost of applying the elementary triangular factors at the cost of $O(C_{sp}k^2)$ operations for each cluster, bringing it to a total cost of $O(C_{sp}k^2 n)$. Since the sparsity constant does not grow with problem size, the complexity of both methods grows linearly with $n$, if we assume a bound $k$ on the rank. 

In terms of memory, each cluster stores the orthogonal factor, the redundant diagonal blocks, and the $O(C_{sp})$ blocks of the elementary triangular factor. This yields a total cost of storing the factors to $O(C_{sp}k^2 n)$, indicating that, under the same assumptions, memory requirements also grow linearly with problem size. %

\section{Numerical Experiments}
\ignore{
\xslnote{The 2D covariance matrix seems to work well. What are the other problems to show? Can we list and define them?
\begin{itemize}
    \item Problem 2: IE operator for the Helmholtz, Eqn (9) in H2-construction paper 
    \item Problem 3: some BEM problem?
    \item Problem 4: Updating H2 with a low rank product
    \item Problem 5: frontal matrices
    \item ...  
\end{itemize}
}  }

We illustrate the algorithm developed herein on four families of symmetric positive definite systems. To construct the initial hierarchical matrices in the following examples, we first generate a uniform grid of points in a $d$-dimensional space and build a cluster tree using a KD-tree with a leaf size $m$. A dual tree traversal on this tree with an admissibility parameter $\eta$ is used to determine the structure of the $\mathcal{H}^2$-matrix. Then the coupling, basis and transfer matrices are generated using Chebyshev interpolation \cite{borm07} of order $p$, where a $p^d$ tensor product grids of Chebyshev interpolation points are overlaid on the bounding boxes of each cluster. The order is increased from an initial order $p_0$ at the leaf level with every other level to compensate for the coarser matrix structures. Algebraic compression is carried out to a specified tolerance $\epsilon$ to reduce the original ranks $k=p^d$ and orthogonalize the basis of the matrix. To help moderate the condition number of the matrices in these applications, we add a small diagonal regularization term $\alpha_r I$. Finally, factorization is carried out to the threshold $\epsilon_{lu}$. The problem sizes vary from $2^{14}$ to $2^{20}$ and we list the parameters for the construction and factorization of the applications in Table \ref{tab:application_params}. The table also shows the maximum sparsity constant $C_{sp}$ of the inadmissible nodes for each problem across all levels for each problem as well as the maximum rank after algebraic compression $k_{max}$, before factorization commences.

\begin{table}[t]
\centering
\begin{tabular}{|c|c|c|c|c|c|c|c||c|c|}
\hline
Problem & $m$ & $p_0$ & $d$ & $\eta$ & $\alpha_r$ & $\epsilon$ & $\epsilon_{lu}$ & $k_{max}$ & $C_{sp}$ \\
\hline
2D Covariance & 64 & 8 & 2 & 0.9 & $10^{-2}$  & $10^{-7}$  & $10^{-6}$ & 39  & 11 \\
\hline
3D Covariance & 64 & 4 & 3 & 0.7  & $10^{-2}$ & $10^{-7}$ & $10^{-6}$ & 111 & 77 \\
\hline
2D Laplace IE & 64 & 8 & 2 & 0.9 & $10^{-5}$ & $10^{-7}$ & $10^{-6}$ & 23 & 11 \\
\hline
3D Helmholtz IE & 64 & 4 & 3 & 0.7 & $10^{-2}$ & $10^{-7}$ & $10^{-6}$ & 86 & 77 \\
\hline
LRU 3D Covariance & 128 & 4 & 3 & 0.9 & $10^{-2}$ & $10^{-8}$ & $10^{-7}$ & 190 & 39 \\
\hline
\end{tabular}
\caption{Parameters for construction and factorization of the problems (leaf size $m$, leaf-level Chebyshev interpolation order $p_0$, physical domain dimension $d$, admissibility constant $\eta$ {\rm [Eq.~(\ref{eqn:admissibility})]}, diagonal regularization term $\alpha_r$, algebraic compression tolerance $\epsilon$, and LU factorization tolerance $\epsilon_{lu}$) and maximum ranks of the initial matrix $k_{max}$ and sparsity constant $C_{sp}$. 
}
\label{tab:application_params}
\end{table}

\subsection{Test Problems}\label{sec:test_problems}
We consider four applications. The first application involves spatial statistics covariance matrices from the Gaussian Process with
the exponential kernel
$  K(x, y) = e^{-\frac{|x - y|}{l}}\label{eq:cov} .$
We test both 2D and 3D problems using a uniform distribution of points in a unit square or cube with correlations lengths $l$ set to $0.1$ and $0.2$, respectively. 
For the second application, we consider the hierarchical matrix constructed in the same way for the free-space Green’s function associated with the 2D Laplace equation, solving the volume integral equation (IE) whose operator is
\begin{equation*}
  K(x, y) = -\frac{1}{2\pi}\log(||x - y||),~ x\neq y\label{eq:lapace2dkernel}.
\end{equation*}

The next two applications construct the hierarchical matrix using adaptive sketching-based construction~\cite{boukaram2025}, where matrix entry evaluation and fast matrix-vector products are all that is needed to construct the hierarchical matrix to a specified error threshold. First, we consider the discretization of a 3D volume integral equation operator for the Helmholtz equation also on a uniform cubic lattice.  The oscillatory IE kernel is  
\begin{equation*}
K(x, y) = \frac{\cos{(\kappa|x - y|)}}{|x - y|},~ x\neq y\label{eq:ie}
\end{equation*}
with wavenumber $\kappa$ fixed at be 3 as the mesh resolution is varied.
The final application involves updating a hierarchical matrix with a global low rank update (LRU). This is commonly encountered during the LU decomposition of hierarchical matrices or in the multifrontal factorization of sparse matrices.
For that, we take a previously generated 3D covariance matrix and apply a random rank $32$ update. We also increase the accuracy for this problem and increased the leaf size to $128$ due to the larger ranks at the leaves making it inefficient to use the same leaf size of $64$ as the other problems. 
\deknote{These examples, typical of many applications, occupy a ``sweet spot'' in which the block ranks do not grow substantially during the factorization. In more general cases, for truly ``blackbox'' factorization software, some intermediate recompression may be required.}






\subsection{Testing Setup}
All tests were run on an AMD Ryzen Threadripper PRO 5995WX 64 core processor system with 512 GB of memory. The OpenMP number of threads is set to 16, which is the point where memory bandwidth becomes the performance limiter for the batched linear algebra. We used the OpenBLAS linear algebra library and set the number of threads used to 1, since most operations are sufficiently small. 

\subsection{Results}
\begin{figure}[t]
    \centering
    \subfigure[Log-log plot of the factorization time for various problem sizes demonstrating the linear runtime of the factorization algorithm.\label{fig:hlu_time}]{\includegraphics[width=\singlecolumnimgwidth]{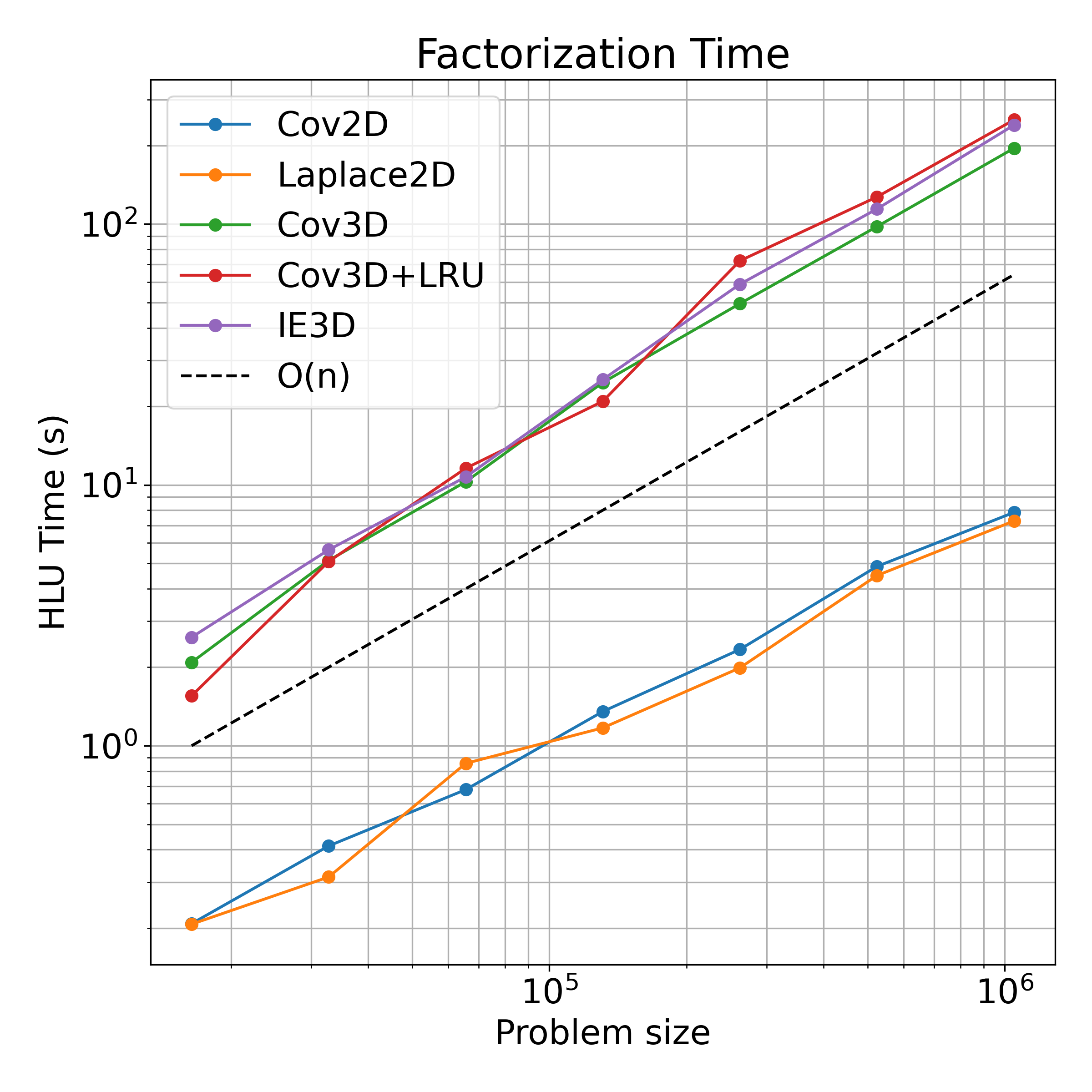}}
    \hfill
    \subfigure[Log-log plot of the memory consumed by the factorization for various problem sizes demonstrating the linear growth in memory.\label{fig:hlu_memory}]{\includegraphics[width=\singlecolumnimgwidth]{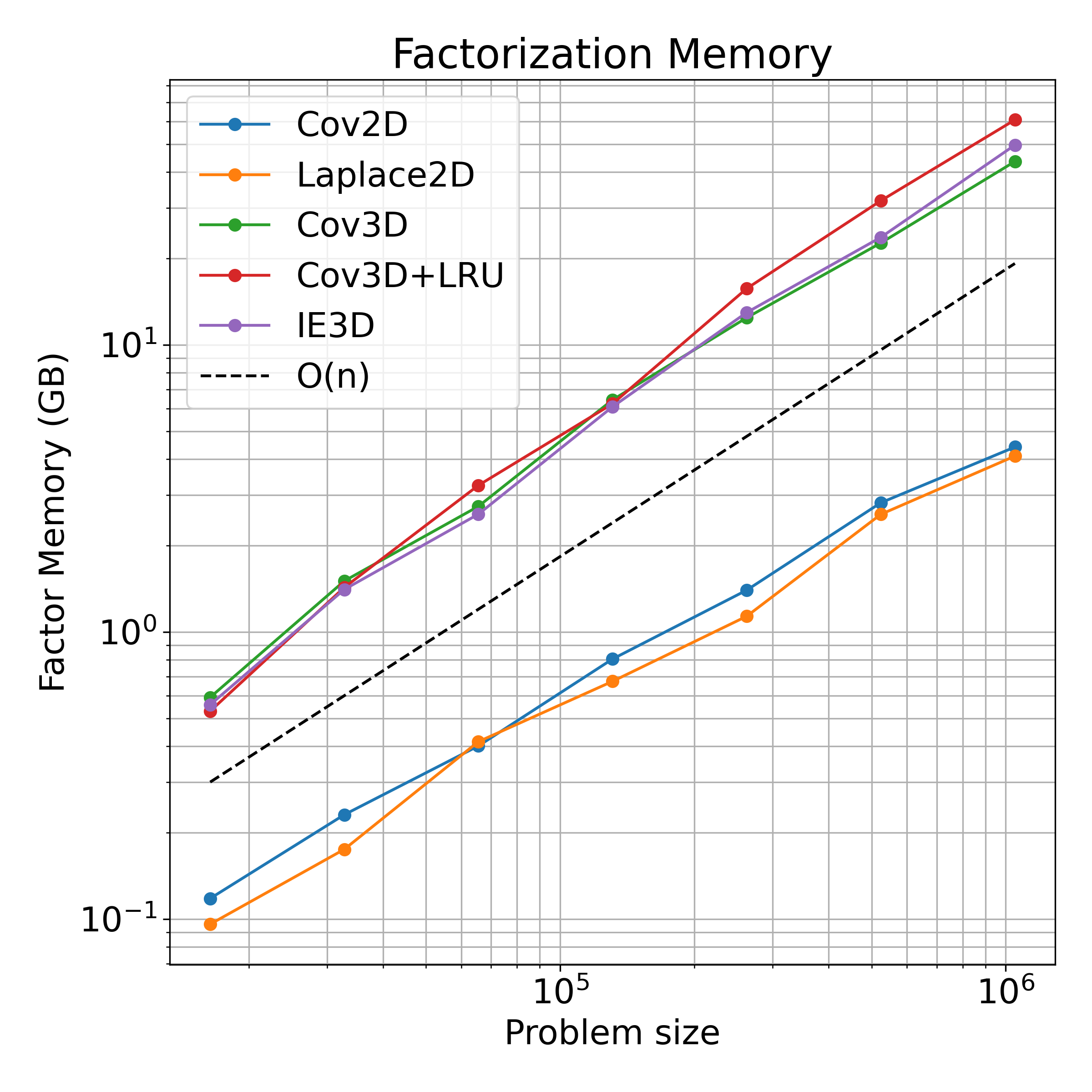}}
    \caption{The factorization algorithm exhibits linear complexity in time and memory.}
    \label{fig:hlu_time_and_memory}
\end{figure}
\begin{figure}[hb]
  \centering
  \includegraphics[width=\doublecolumnimgwidth]{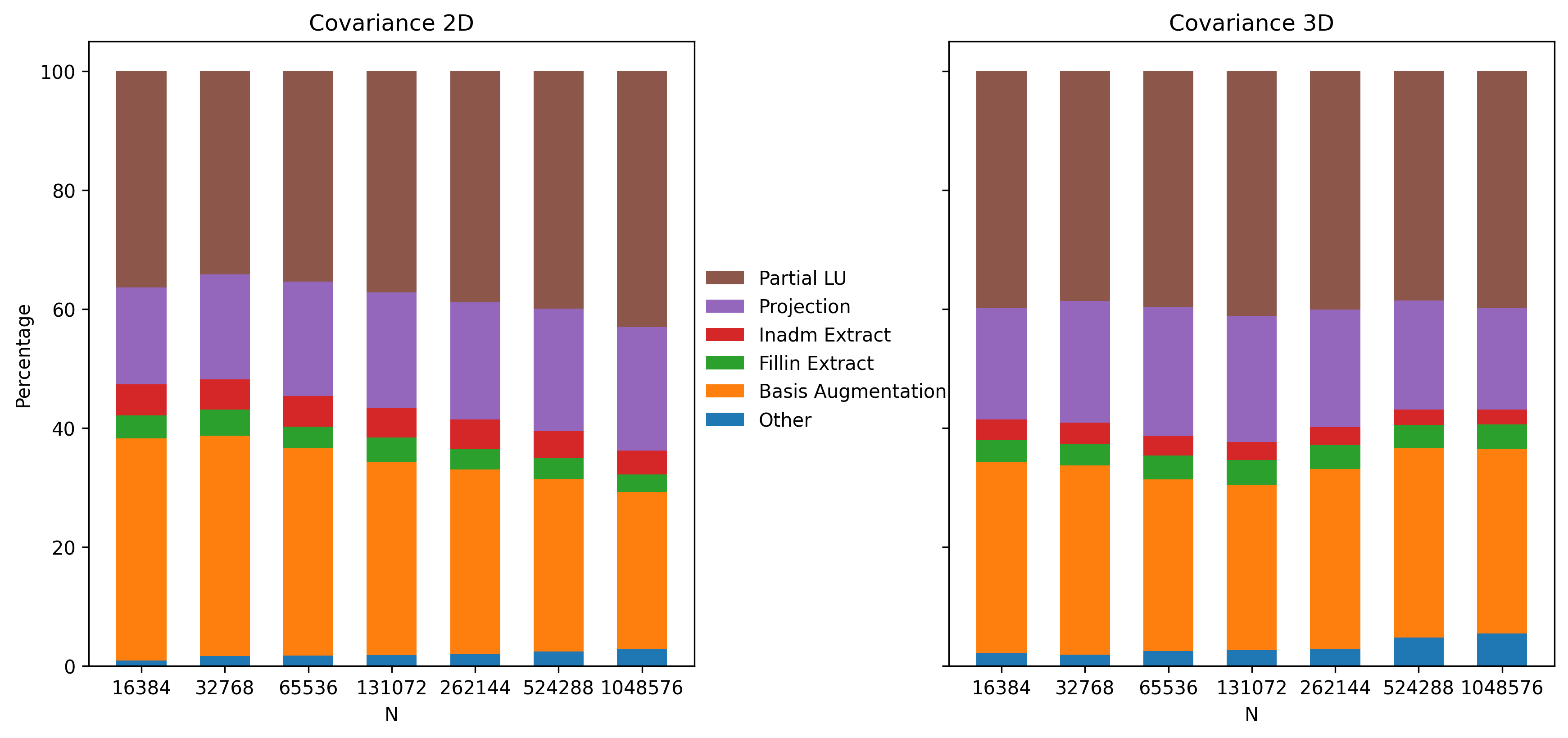}  
  \caption{The percentage of time spent in each of the major computational kernels of the factorization for the 2D (left) and 3D (right) covariance matrices.}
  \label{fig:profile}
\end{figure}

\paragraph{Factorization.}
Figure~\ref{fig:hlu_time} shows the factorization times for our applications along with  $O(n)$ reference lines,
clearly demonstrating the linear complexity of the algorithm. 
The 3D problems are considerably slower than the 2D ones due not only to higher ranks,
but also because the sparsity constant tends to be higher for 3D problems, as can be seen in Table \ref{tab:application_params},
due to the $O(C_{sp}^2k^3 n)$ complexity of the algorithm.  Figure~\ref{fig:profile} breaks down the runtime of the factorization into its major phases, with the partial LU dominating runtime as expected. 
The basis augmentation step, where the QR decomposition of the fill-in block row followed by an SVD, is a close second despite being $O(C_{sp})$ fewer operations. This is partially due to the fact that the matrix multiplications in the Schur complement updates are far more efficient operations, and partially due to the seemingly poor scaling of the QR decompositions of the basis augmentation with the number of threads. This seems to be the case for both the 2D and 3D problems despite the significantly larger sparsity constants of the 3D problem. 
\begin{table}[ht]
\centering
\begin{tabular}{rrrrrrrrr}
\toprule
Threads & HLU 2D & Solve 2D & HLU 3D & Solve 3D & GEMM (S) & GEMM (L) & QR (S) & QR (L) \\
\midrule
1 & 28.280 & 0.376 & 848.640 & 2.389 & 1.850 & 46.470 & 37.580 & 1141.540 \\
2 & 15.020 & 0.222 & 447.573 & 1.448 & 1.220 & 23.950 & 23.370 & 582.340 \\
4 & 8.812 & 0.173 & 258.186 & 1.214 & 0.970 & 12.440 & 12.560 & 299.510 \\
8 & 7.099 & 0.172 & 204.928 & 1.202 & 1.140 & 7.330 & 17.570 & 179.650 \\
16 & 7.881 & 0.133 & 142.617 & 0.768 & 1.170 & 4.400 & 36.490 & 198.180 \\
\bottomrule
\end{tabular}
\caption{Thread scaling of the factorization and solve (in seconds) for the 2D and 3D covariance matrices, and the OpenMP parallelized batched GEMM and QR operations for a batch of 1000 small and large matrix operands (in milliseconds). 
}
\label{tab:scaling}
\end{table}
To more clearly illustrate this, Table \ref{tab:scaling} shows how increasing the number of OpenMP threads changes the factorization and solve times (in seconds) for the $n=2^{20}$ 2D and 3D covariance matrices and the time (in milliseconds) for 1000 operations for GEMM and QR batches for small (S) and large (L) operands. For the GEMMs, small operands are $30\times 30$ and large operand dimensions are $100 \times 100$, while the QR small operands are $300 \times 30$ and large operands are $1000 \times 100$. Clearly, adjusting the number of threads used in each batched routine based on operand sizes would lead to more efficient batched routines like those specialized on the GPU and could yield some performance benefits, which is future work.

\begin{figure}[hb]
  \centering
  \includegraphics[width=\doublecolumnimgwidth]{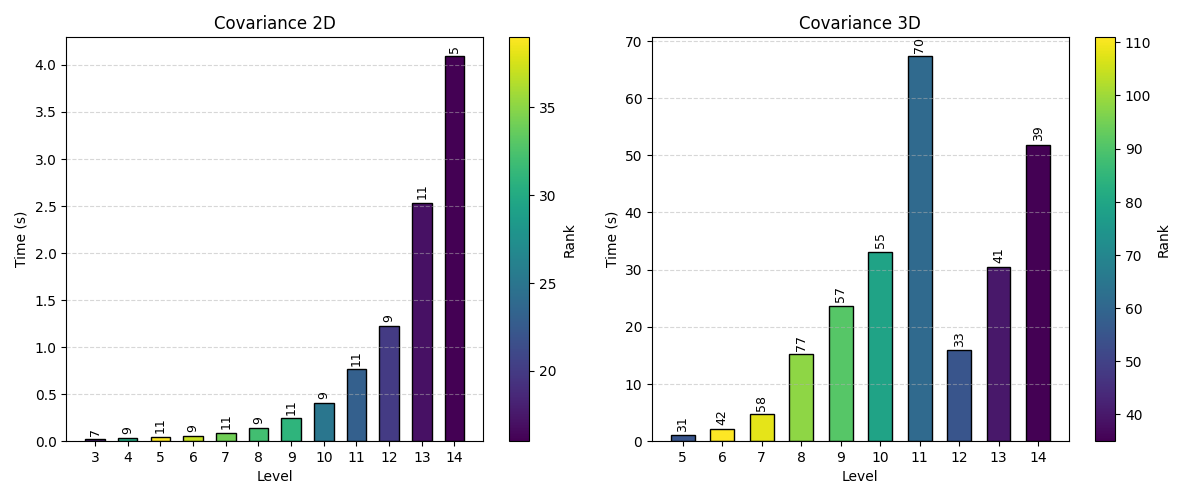}  
  \caption{The time spent in each level of the factorization for $N=2^{20}$ for the 2D (left) and 3D (right) covariance matrices. The sparsity constant of the level is labeled on top of each bar and the maximum rank of each level is color coded into the bars.}
  \label{fig:per_level_times}
\end{figure}

Replacing the Householder QR decompositions with Cholesky QR could trade accuracy for greater efficiency in the basis augmentation. Alternately, the Gram matrix approach followed by an eigenvalue decomposition could be preferable. The performance of the 2D problems is mostly bandwidth limited due to the smaller ranks, limiting the benefits of increasing the number of threads used in the batched routines. 
The smaller batch sizes generated in the smaller problems also prevent threads from being fully utilized. Figure \ref{fig:per_level_times} shows the time taken at each level of the factorization for the 2D and 3D covariance matrices. The sudden jump in the sparsity constant for the 3D problem at level 11 is due to the nonisotropic distribution of the points in the $128 \times 128 \times 64$ ($=2^{20}$ discrete size) 3D cube. This increase along with the larger ranks at that level makes the upper level take more time than the leaf level despite having only $1/8$ of the clusters, showing that the leaf level is not necessarily the most expensive part of the computation. 
The expected linear growth of the memory consumed by the factors of the matrix is also clearly shown in Figure~\ref{fig:hlu_memory}. The 3D problems again consume more memory than the 2D problems due to the aforementioned differences in rank and sparsity. 

\begin{figure}[!htb]
    \centering
    \subfigure[Log-log plot of the solve time for various problem sizes demonstrating the linear runtime of the solve algorithm.\label{fig:hlu_solve}]{\includegraphics[width=\singlecolumnimgwidth]{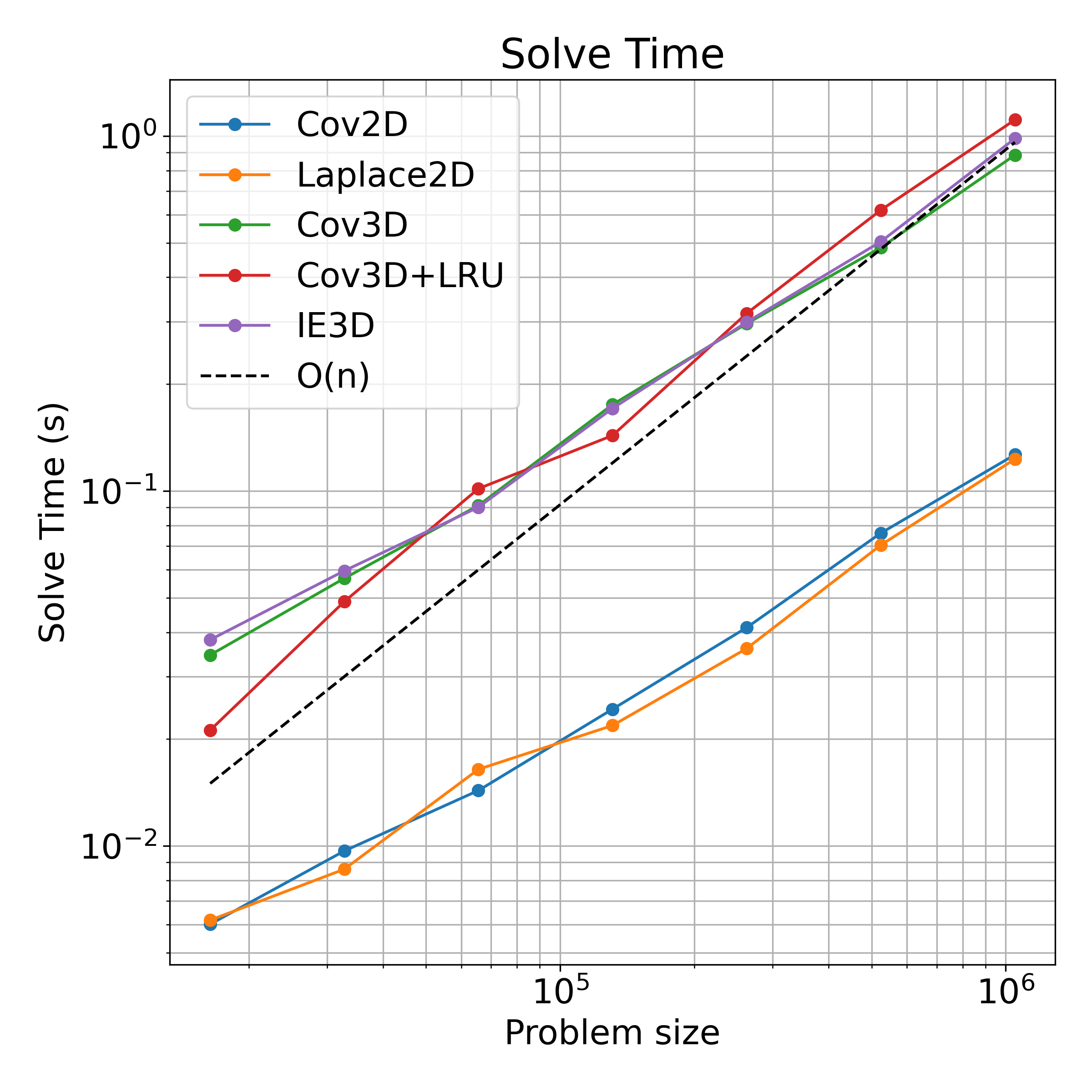}}
    \hfill
    \subfigure[Log-log plot of the backward error of the factorization.\label{fig:hlu_error}]{\includegraphics[width=\singlecolumnimgwidth]{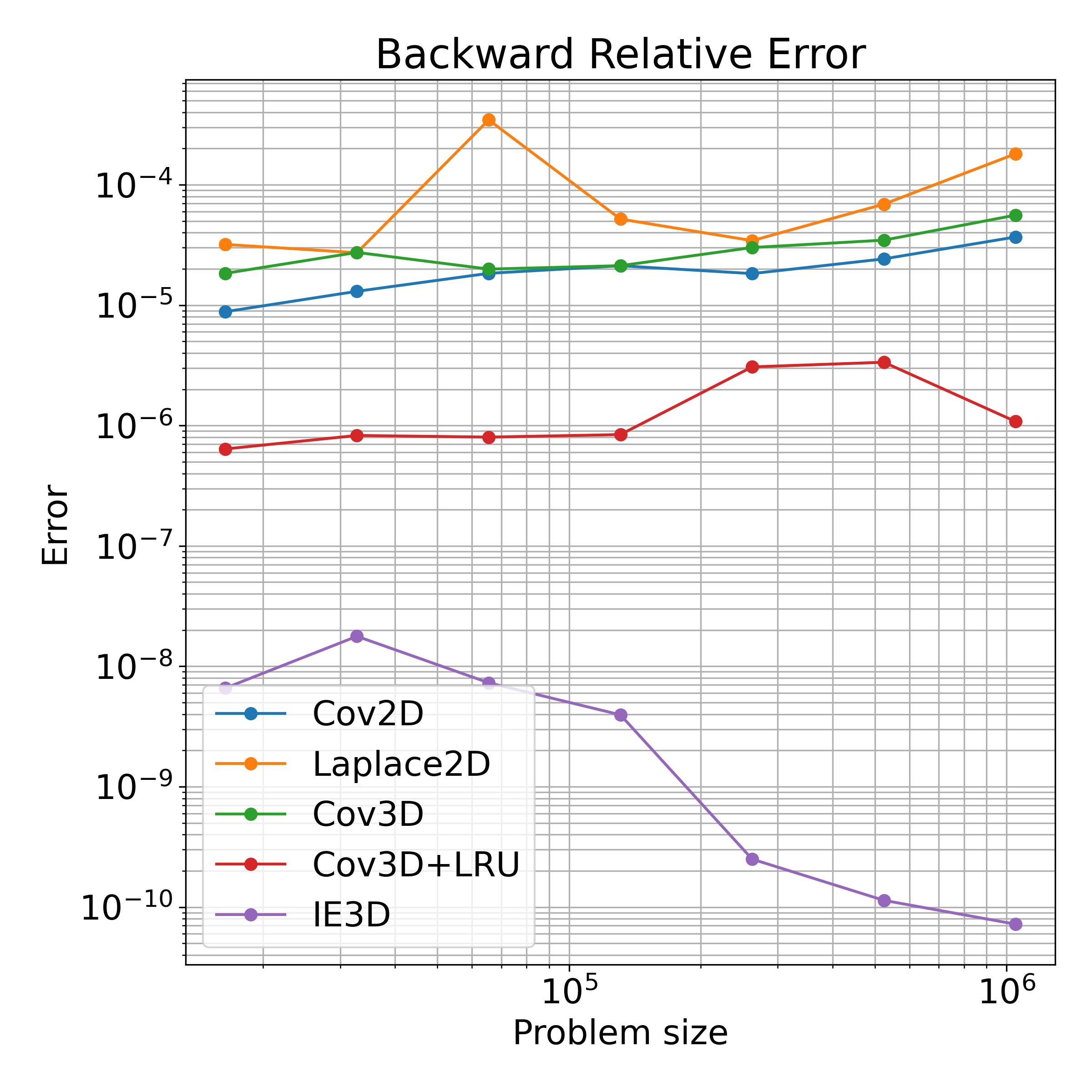}}
    \caption{The factorization algorithm exhibits linear complexity in solve time and controls the error well.}
    \label{fig:solvetime_and_res_error}
\end{figure}

\paragraph{Solve.}
Figure~\ref{fig:hlu_solve} shows the solve times for our applications for a randomly generated right hand side vector $b$ along with the $O(n)$ reference lines, showing its overall linear growth. Unlike the factorization, all operations are bandwidth limited in the solve phase and the overhead of the smaller batch sizes have a clearer impact on runtime for all applications. To compute the error in the factorization, we generate a random vector $x$, multiply it by the matrix $A$ to generate the right hand side $b = Ax$, and solve for $\tilde{x} = A^{-1} b$. The relative backward error is computed as $e_b = ||A\tilde{x} - b|| / ||b||$ and is shown in Figure \ref{fig:hlu_error}, demonstrating the high accuracy achieved by the factorization for the given thresholds.

\section{Conclusion}

Building on prior work for $\mathcal{H}^2$ matrix construction on GPUs for classes of matrices possessing data sparsity, which arise broadly in contemporary large-scale applications, we \deknote{provide new  ``blackbox''} factorization and solution routines exploiting the same recursive skeletonization structure and the similar modular composition. Demonstrations herein include covariance matrices in 2D and 3D, a 2D integral equation with a monotonic kernel, a 3D integral equation with an oscillatory kernel, and a global low-rank update to a covariance matrix. We demonstrate the expected linear complexity in execution time and memory for systems up to one million in size, using a multi-threaded CPU implementation.  

To promote a culture of reproducibility, full parameter tunings for the computational experiments are documented herein and we are releasing code for the solver in open-source. The constants of the linear scaling are described theoretically and coefficient values documented in the experiments. Algorithmic decisions for various modules are described and alternatives listed. Memory-bound phases of the algorithm are noted. At the time of submission, a GPU implementation is in progress.

We anticipate a steady appetite for exploiting data sparsity in practical large-scale applications in modeling and simulation, data analytics, and machine learning on which hierarchically low-rank solvers such as the one described herein will be deployed.

\bigskip
{\em The authors acknowledge their deep and several debts to Nicholas Higham in many aspects of their careers and work: research, pedagogy, and personal inspiration.  They dedicate this particular body of work to his profound legacy in computational linear algebra and far, far beyond.}

\section*{Acknowledgments.}

This work was supported in part by the U.S. Department of Energy, Office of Science, Office of Advanced Scientific Computing Research's Applied Mathematics Competitive Portfolios program and Scientific
Discovery through Advanced Computing (SciDAC) program
through the FASTMath Institute under Contract No. DE-AC02-05CH11231.




\bibliographystyle{plain}
\bibliography{H-matrices}

\begin{thebibliography}{10}

\bibitem{ambikasaran2013mathcal}
Sivaram Ambikasaran and Eric Darve.
\newblock {An $\mathcal{O}(N \log{N})$ fast direct solver for partial
  hierarchically semi-separable matrices: with application to radial basis
  function interpolation}.
\newblock {\em Journal of Scientific Computing}, 57:477--501, 2013.

\bibitem{ambikasaran2015fast}
Sivaram Ambikasaran, Daniel Foreman-Mackey, Leslie Greengard, David~W Hogg, and
  Michael O’Neil.
\newblock {Fast direct methods for Gaussian processes}.
\newblock {\em IEEE Transactions on Pattern Analysis and Machine Intelligence},
  38(2):252--265, 2015.

\bibitem{bebendorf2003existence}
Mario Bebendorf and Wolfgang Hackbusch.
\newblock {Existence of $\mathcal{H}$-matrix approximants to the inverse
  FE-matrix of elliptic operators with $L^{\infty}$-coefficients}.
\newblock {\em Numerische Mathematik}, 95:1--28, 2003.

\bibitem{borm07}
Steffen B{\"o}rm and Jochen Garcke.
\newblock Approximating {G}aussian processes with {$\mathcal{H}^2$}-matrices.
\newblock In {\em European Conference on Machine Learning}, pages 42--53.
  Springer, 2007.

\bibitem{borm2003introduction}
Steffen B{\"o}rm, Lars Grasedyck, and Wolfgang Hackbusch.
\newblock Introduction to hierarchical matrices with applications.
\newblock {\em Engineering analysis with boundary elements}, 27(5):405--422,
  2003.

\bibitem{borm13_lr}
Steffen B{\"o}rm and Kerstin Reimer.
\newblock Efficient arithmetic operations for rank-structured matrices based on
  hierarchical low-rank updates.
\newblock {\em Computing and Visualization in Science}, 16:247--258, 2013.

\bibitem{boukaram2025}
Wajih Boukaram, Yang Liu, Pieter Ghysels, and Xiaoye~Sherry Li.
\newblock {Adaptive Sketching Based Construction of H2 Matrices on GPUs}.
\newblock In {\em The 26th IEEE International Workshop on Parallel and
  Distributed Scientific and Engineering Computing (PDSEC 2025)}. IEEE, 06
  2025.
\newblock Best Paper Award.

\bibitem{boukaram19_gpu}
Wajih~Halim Boukaram, George Turkiyyah, and David Keyes.
\newblock Hierarchical matrix operations on {GPU}s: Matrix-vector
  multiplication and compression.
\newblock {\em ACM Transactions on Mathematical Software}, 45(1):1--28,
  February 2019.

\bibitem{chandrasekaran2007fast}
Shiv Chandrasekaran, Patrick Dewilde, Ming Gu, William Lyons, and Timothy Pals.
\newblock {A fast solver for HSS representations via sparse matrices}.
\newblock {\em SIAM Journal on Matrix Analysis and Applications}, 29(1):67--81,
  2007.

\bibitem{chandrasekaran2008superfast}
Shiv Chandrasekaran, Ming Gu, Xiaotian Sun, Jianlin Xia, and Jiang Zhu.
\newblock {A superfast algorithm for Toeplitz systems of linear equations}.
\newblock {\em SIAM Journal on Matrix Analysis and Applications},
  29(4):1247--1266, 2008.

\bibitem{chavez2020scalable}
Gustavo Ch{\'a}vez, Yang Liu, Pieter Ghysels, Xiaoye~Sherry Li, and Elizaveta
  Rebrova.
\newblock Scalable and memory-efficient kernel ridge regression.
\newblock In {\em 2020 IEEE International Parallel and Distributed Processing
  Symposium (IPDPS)}, pages 956--965. IEEE, 2020.

\bibitem{coulier2017inverse}
Pieter Coulier, Hadi Pouransari, and Eric Darve.
\newblock The inverse fast multipole method: using a fast approximate direct
  solver as a preconditioner for dense linear systems.
\newblock {\em SIAM Journal on Scientific Computing}, 39(3):A761--A796, 2017.

\bibitem{ghysels2017robust}
Pieter Ghysels, Sherry~Li Xiaoye, Christopher Gorman, and Fran{\c{c}}ois-Henry
  Rouet.
\newblock A robust parallel preconditioner for indefinite systems using
  hierarchical matrices and randomized sampling.
\newblock In {\em 2017 IEEE International Parallel and Distributed Processing
  Symposium (IPDPS)}, pages 897--906. IEEE, 2017.

\bibitem{gillman2012direct}
Adrianna Gillman, Patrick~M Young, and Per-Gunnar Martinsson.
\newblock {A direct solver with $O(N)$ complexity for integral equations on
  one-dimensional domains}.
\newblock {\em Frontiers of Mathematics in China}, 7:217--247, 2012.

\bibitem{hackbusch1999sparse}
Wolfgang Hackbusch.
\newblock {A sparse matrix arithmetic based on $\mathcal{H}$-matrices. Part I:
  Introduction to-matrices}.
\newblock {\em Computing}, 62(2):89--108, 1999.

\bibitem{keyes20_haha}
David Keyes, Hatem Ltaief, and George Turkiyyah.
\newblock {Hierarchical algorithms on hierarchical architectures}.
\newblock {\em Philosophical Transactions of the Royal Society A: Mathematical,
  Physical and Engineering Sciences}, 378(2166):20190055, January 2020.

\bibitem{l2016hierarchical}
Kenneth L.~Ho and Lexing Ying.
\newblock {Hierarchical interpolative factorization for elliptic operators:
  integral equations}.
\newblock {\em Communications on Pure and Applied Mathematics},
  69(7):1314--1353, 2016.

\bibitem{liang2024on}
Tianyu Liang, Chao Chen, Per-Gunnar Martinsson, and George Biros.
\newblock {An $O(N)$ distributed-memory parallel direct solver for planar
  integral equations}.
\newblock In {\em 2024 IEEE International Parallel and Distributed Processing
  Symposium (IPDPS)}, pages 440--452. IEEE, 2024.

\bibitem{Ma2019}
Miaomiao Ma and Dan Jiao.
\newblock {Direct solution of general ${\mathcal{H}}^2$ -matrices with
  controlled accuracy and concurrent change of cluster bases for
  electromagnetic analysis}.
\newblock {\em IEEE Transactions on Microwave Theory and Techniques},
  67(6):2114--2127, 2019.

\bibitem{ma2022scalable}
Qianxiang Ma, Sameer Deshmukh, and Rio Yokota.
\newblock {Scalable linear time dense direct solver for 3-d problems without
  trailing sub-matrix dependencies}.
\newblock In {\em SC22: International Conference for High Performance
  Computing, Networking, Storage and Analysis}, pages 1--12. IEEE, 2022.

\bibitem{Ma_2024}
Qianxiang Ma and Rio Yokota.
\newblock {An inherently parallel \hh-ULV factorization for solving dense
  linear systems on GPUs}.
\newblock {\em The International Journal of High Performance Computing
  Applications}, 38(4):314–336, April 2024.

\bibitem{martinsson05}
Per-Gunnar Martinsson and Vladimir Rokhlin.
\newblock A fast direct solver for boundary integral equations in two
  dimensions.
\newblock {\em Journal of Computational Physics}, 205(1):1--23, 2005.

\bibitem{minden2017recursive}
Victor Minden, Kenneth~L Ho, Anil Damle, and Lexing Ying.
\newblock A recursive skeletonization factorization based on strong
  admissibility.
\newblock {\em Multiscale Modeling \& Simulation}, 15(2):768--796, 2017.

\bibitem{Sushnikova2023}
Daria Sushnikova, Leslie Greengard, Michael O’Neil, and Manas Rachh.
\newblock {FMM-LU: A Fast Direct Solver for Multiscale Boundary Integral
  Equations in Three Dimensions}.
\newblock {\em Multiscale Modeling \& Simulation}, 21(4):1570--1601, 2023.

\bibitem{xia2013randomized}
Jianlin Xia.
\newblock Randomized sparse direct solvers.
\newblock {\em SIAM Journal on Matrix Analysis and Applications},
  34(1):197--227, 2013.

\bibitem{yesypenko2025}
Anna Yesypenko and Per-Gunnar Martinsson.
\newblock {Randomized Strong Recursive Skeletonization: Simultaneous
  Compression and LU Factorization of Hierarchical Matrices using Matrix-Vector
  Products}, 2025.

\bibitem{h2opus}
Stefano Zampini, Wajih Boukaram, George Turkiyyah, Omar Knio, and David Keyes.
\newblock {H2Opus:} a distributed-memory multi-{GPU} software package for
  non-local operators.
\newblock {\em Advances in Computational Mathematics}, 48(3), May 2022.

\end{thebibliography}


\end{document}